\documentclass{jfm}

\usepackage{amsmath}
\usepackage{amssymb}
\usepackage{mathtools}
\usepackage{graphicx}
\usepackage[dvipsnames]{xcolor}
\usepackage{adjustbox}
\usepackage{multirow}
\usepackage{dcolumn}
\usepackage{bm}
\usepackage{epstopdf, epsfig}
\usepackage{subfig}
\usepackage{tikz}
\usepackage[normalem]{ulem}

\renewcommand\Rey{\mbox{\textit{Re}}}  
\newcommand\Sc{\mbox{\textit{Sc}}} 
\newcommand\Fr{\mbox{\textit{Fr}}}  
\newcommand\Ri{\mbox{\textit{Ri}}}  
\newcommand{\be}[1]{\begin{equation}\label{#1}}
\newcommand{\ee}{\end{equation}}
\newcommand{\ba}[1]{\begin{eqnarray}\label{#1}}
\newcommand{\ea}{\end{eqnarray}}
\newcommand{\rf}[1]{(\ref{#1})}
\newcommand{\nn}{\nonumber}
\newcommand{\dd}{{\rm{d}}}
\newcommand{\im}{{\rm{i}}}

\shorttitle{Localized layers in stratified Poiseuille flow}
\shortauthor{J. Labarbe, P. Le Gal, U. Harlander, S. Le Diz\`es and B. Favier}

\title{Localized layers of turbulence in stratified horizontally sheared Poiseuille flow}

\author{
  J. Labarbe\aff{1}
  \corresp{\email{joris.LABARBE@univ-amu.fr}},
  P. Le Gal\aff{1},
  U. Harlander\aff{2},
  S. Le Diz\`es\aff{1} \and
  B. Favier\aff{1}
  }

\affiliation{
\aff{1}Aix Marseille Univ., CNRS, Centrale Marseille, IRPHE, Marseille, France
\aff{2}Department of Aerodynamics and Fluid Mechanics, Brandenburg University of Technology, Cottbus-Senftenberg, Germany
}

\begin{document}

\maketitle

\begin{abstract}
This article presents a numerical analysis of the instability developing {\color{black}{in horizontally sheared Poiseuille flow, when stratification extends along the vertical direction}}. Our study builds up on the previous work that originally detected the linear instability of such configuration, by means of experiments, theoretical analysis and numerical simulations \citep{G21}. We extend hereafter this former investigation beyond linear theory, investigating nonlinear regimes with direct numerical simulations. {\color{black}{We notice that the flow loses its vertical homogeneity through a secondary bifurcation}}, due to harmonic resonances, and further describe this symmetry-breaking mechanism in the vicinity of the instability threshold. When departing away from this limit, we observe a series of bursting events that break down the flow into disordered motions driven by localized shear instabilities. This intermittent dynamics leads to the coexistence of horizontal localized layers of stratified turbulence surrounded by quiescent regions of meandering waves.
\end{abstract}


\section{Introduction}

Turbulent stratified flows are ubiquitous in nature, most notably in the Earth atmosphere or in oceans, but are highly challenging to study in laboratories. This impediment is due to the requirement of very large installations to monitor the different characteristic scales of turbulence. It is then of interest to construct idealized hydrodynamical models to more easily capture the transition towards disordered motions and explore connections with the spontaneous formation of density layers in stratified fluids \citep{OCW13}. Indeed, the challenge of clearly explaining the layering and mixing, naturally present in geophysical flows, remains nowadays still open (see, e.g., \cite{C21} for a recent review). This question has fundamental ramifications as it directly relates to the transport of heat, pollutants or bio-mass in the density-stratified fluids present on Earth. Over the past decades, several contributions to the study of turbulent stratified flows have been made, seeking for a complete description of the mechanisms responsible for layering and mixing \citep{TDCK16,ZTC17a,ZTC17b}. These works notably demonstrated that stratified turbulence inherits a strong anisotropy while developing and that inhomogeneous diffusive processes, such as mixing, are a direct consequence of this spatially intermittent phenomenon. 

The novelty of our approach lies in the generation of turbulence through instabilities of a model flow, namely the vertically-stratified horizontal plane Poiseuille flow \citep{G21}. It is noteworthy to emphasize the difference with the more classical bi-dimensional case, where both stratification and shear lie on the same plane \citep{GR68}. In fact, it is known since the studies of \cite{BT84} and \cite{BF09a,BF09b} that most environmental flows spontaneously generate internal gravity waves, whenever the stratification is oriented perpendicularly to the direction of motion. In addition, these small-scale waves propagate (with a Doppler-shift) and interact within the density layers, leading to instabilities \citep{S81}. Historically, wave resonances in parallel sheared fluids have been introduced in the seminal paper of \cite{C79} and further extended to diverse configurations of stratified flows (see, for instance, the review by \cite{C11} on this topic). Moreover, such instabilities can transit to turbulence due to the collapse of finite disturbances, once the system saturates and reaches the nonlinear regime \citep{C94}. One benefit in considering instabilities is therefore the unnecessary need to trigger turbulence by explicit forcing, since the dynamics is self-sustained by definition. Hence, our study makes use of these instabilities to follow the route towards stratified turbulence, as done in a plethora of recent investigations on linear instabilities of stratified flows. We notably rely on the observations made for Taylor--Couette flows \citep{MMY01,BG07,DX10,PB13}, plane Couette flows with spanwise stratification \citep{Couette18,LCK19}, and a selection of rotating flows \citep{BC00,DB09,RDM11}. All the latter instabilities are indeed caused by the resonant interaction of Doppler-shifted internal gravity waves, which is the case in our context as well \citep{G21}.

In this article, we demonstrate the presence of a symmetry-breaking mechanism, when the primary linear instability saturates and the flow enters the nonlinear regime, with the appearance of a spatial modulation of the basic plane Poiseuille flow profile. This modification in the streamwise mean velocity profile is a direct consequence of the {\color{black}{nonlinear}} interaction of harmonics, originally generated by counter-propagating internal gravity waves. The loss of invariance along the vertical additionally results in the formation of a localized region where the velocity fluctuates, inducing new spanwise shear in the system. 
{\color{black}{This process is quite original and differs from the transition scenario of the unstratified plane Poiseuille flow as described in the review article of \cite{Laurette}. The non linear regime and the transition to turbulence of the unstratified plane Poiseuille flow, often called the channel flow, manifests in the form of turbulent bands which are inclined with respect to the base flow and possess a
wavelength ranging from 20 to 40 times the half gap of the channel \citep{Laurette}. The first
observations of these bands in the plane Poiseuille flow came from the numerical simulations of
\cite{Tsukahara}. They were followed by the experiments of \cite{Hashimoto}. These first descriptions were then completed more recently by numerical
simulations \citep{Laurette2,Kashyap} where it is shown in particular that when decreasing the Reynolds
number, these turbulent bands appear from homogeneous turbulence through a continuous
modulation process of the turbulent fields. As demonstrated by \cite{Reetz}, the link
between these inclined turbulent bands and exact invariant solutions of the Navier-Stokes equation is
not trivial.}} 

While the vertical gradients are sharpened, the flow eventually becomes subject to secondary instabilities that further saturate and break down into turbulence, through a series of bursting events \citep{LCK17}. We describe these spatially localized layers of turbulence by means of direct numerical simulations (DNS) in a doubly-periodic geometry with finite extension in the wall-normal direction. We support our observations with the computation of local measures of vertical shear from our direct computations to highlight this novel phenomenology.

The paper is structured as follows. Section \ref{Sec2} introduces the mathematical settings. Section \ref{Sec3} presents the linear stability analysis of our configuration and how the stability of discrete harmonics is determined accordingly. Section \ref{Sec4} is dedicated to DNS close to the instability threshold. We further discuss on the symmetry-breaking mechanism and the mean-flow interaction it induces. Section \ref{Sec6} describes the phenomenology observed when departing away from the onset of instability. We observe here the triggering of stratified turbulence, initiated from the mechanisms reported previously. Finally, we conclude our study in section \ref{Sec7} with some discussions and future extensions of this work.

\section{Mathematical formulation}
\label{Sec2}

The present work deals with a model of incompressible linearly stratified shear flow enclosed between two parallel walls a distance $D$ apart \citep{G21}. This stratification is assumed to be directed along the orthogonal axis to that of the shear plane. The geometry considered here is the Cartesian frame of reference $(\bm{e}_X,\bm{e}_Y,\bm{e}_Z)$, with corresponding coordinates $\bm{X}=(X,Y,Z)$, describing the streamwise, cross-stream and vertical directions, respectively. {\color{black}{Boundary conditions are no-slip and insulating on both walls at $Y=\pm D/2$. In addition, we assume periodicity along the $(X,Z)$ directions.}}

We express the buoyancy field as
\begin{equation}
\label{rho}
\rho(\bm{X},T) = \rho_L(Z) + \rho'(\bm{X},T) ,
\end{equation}
where $\rho'$ represents the fluctuating contribution and $T$ is the dimensional time. Since we assume a stable stratification, the linear profile in \rf{rho} is written as $\rho_L(Z)=\rho_0(1-ZN^2/g)$, where $N$ is the Brunt-V{\"a}is{\"a}l{\"a} frequency $N=\sqrt{-(g/\rho_0)(\dd\rho_L/\dd Z)}$ (assumed to be real and constant) and $g$ is the constant acceleration due to gravity.

We render this configuration non-dimensional by means of an advective time scale $\tau=D/(2U)$, with $U$ being {\color{black}{the}} local maximum of the mean velocity and {\color{black}{$D/2$ the half gap}}. Coordinates are scaled accordingly, whereas pressure $p$ and buoyancy $b$ are expressed in the units of $\rho_0 U^2$ and $(2\rho_0N^2)/(gD)$, respectively. Hence, the non-dimensional velocity field $\bm{u}=(u,v,w)$, pressure $p$ and buoyancy $b$ are governed by the following set of equations
\begin{subequations}
\label{eom}
\begin{align}
\partial_t \bm{u} + \left( \bm{u} \bm{\cdot} \bm{\nabla} \right) \bm{u} &= - \bm{\nabla} p - \Fr^{-2} b \bm{e}_z + \Rey^{-1} \nabla^2 \bm{u} + f\bm{e}_x , \label{eom1} \\
\partial_t b + \left( \bm{u} \bm{\cdot} \bm{\nabla} \right) b &= w + \left(\Rey\Sc\right)^{-1} \nabla^2 b , \label{eom2} \\
\bm{\nabla} \bm{\cdot} \bm{u} &= 0 , \label{eom3}
\end{align}
\end{subequations}
where $f$ is a spatially uniform term enforcing the streamwise shear profile. This force can either correspond to a steady pressure gradient $f = -2\Rey^{-1}$ or an unsteady function ensuring conservation of the total mass flux $Q = \int_{-1}^{1} \int_{-L_x/2}^{L_x/2} u \dd x \dd y$. In both cases, and in the absence of instabilities, the balance between this external force and viscosity leads to the same base state velocity profile, the well-known plane Poiseuille solution $U_0(y) = 1 - y^2$, that is invariant in the $z$-coordinate. However, if the base system is unstable, we expect the choice of {\color{black}{forcing to
influence the nonlinear regime}}. We emphasize that equations \rf{eom} contain the Reynolds, Froude and Schmidt numbers as control parameters, defined by
\begin{equation}
\Rey = \frac{UD}{2\nu} , \quad \Fr = \frac{2U}{DN} , \quad \Sc = \frac{\nu}{\kappa} ,
\end{equation}
where $\nu$ and $\kappa$ are the constant kinematic viscosity and diffusivity, respectively.

{\color{black}{We perform DNS}} of \rf{eom} by means of the spectral elements solver \textsc{Nek5000} \citep{F97,F07}. Our numerical domain consists in a rectangular parallelepiped of fixed size $\mathcal{D} = [-L_x/2,L_x/2] \times [-1,1] \times [-L_z/2,L_z/2]$, designed such that the most unstable linear mode can develop. In addition, we assume periodic conditions and equispaced elements in both $(x,z)$-directions. We use a non-uniform distribution of wall-normal elements to refine the mesh closer to the walls. {\color{black}{Numerical convergence of the results was checked by increasing the spectral order $N_s$ and comparing the theoretical predictions of growth rates from linear theory with the numerical exponential growth in vertical kinetic energy. Agreement was assumed satisfactory when the absolute difference between the latter was around $\sim 1\%$. We also monitored a few statistical quantities during the saturation phase, such as the viscous dissipation or the vertical kinetic energy (both defined below). Once a fixed number of elements and quadrature nodes was considered sufficient, we still multiplied the total by $1.5$ or $2$ to reach a high-end accuracy. In summary, }}computations were done using $10$ elements per unstable streamwise wavelengths,
$16$ to $24$ elements in the cross-stream direction and $18$ to $24$ elements along the vertical for a total of $\sim 13000$ to $15000$ elements, with a spectral order $N_s\in[8,12]$ (number of Gauss-Lobatto collocations points) and fixed values of $Sc=1$ and $Fr=2$. Following \cite{G21}, the flow is initiated from the parabolic Poiseuille profile plus some random infinitesimal perturbations on the buoyancy field. Since we did not observed noticeable changes in the dynamics of the flow by using a constant pressure gradient or a constant flux, we therefore focused our attention on the latter to better fit with the experimental conditions described in \cite{G21}.

\section{Linear stability analysis}
\label{Sec3}

\subsection{Global stability analysis}

We perform a linear stability analysis of equations \rf{eom} by means of a pseudo-spectral collocation method. Perturbations of the basic state are expressed in terms of normal modes, taking advantage of the periodicity of the flow. Therefore, {\color{black}{we introduce real spatial wavenumbers $(k_x,k_z)$}}, as well as the complex frequency $\omega$, to expand perturbations $\bm{q}'=(\bm{u}',p',b')$ as
\begin{equation}
\label{ansatz}
\bm{q}'(\bm{x},t) = \hat{\bm{q}}(y) \exp \left[ \im ( k_x x + k_z z - \omega t ) \right] + \textrm{c.c.} ,
\end{equation}
where $\textrm{c.c.}$ stands for `complex conjugate'.

Substituting \rf{ansatz} within \rf{eom}, while taking the divergence of \rf{eom1}, we recover a differential expression for the pressure eigenfunction. The latter allows us to express the stratified Orr-Sommerfeld equation, for the cross-stream velocity perturbation $\hat{v}$. Subsequently, we recover the Squire equation, governing the perturbation of axial vorticity $\hat{\eta}$, by applying the curl operator on the wall-normal projection of \rf{eom1}. The resulting linear system yields
\begin{subequations}
\label{leom}
\begin{align}
\left[ k_x \left( U_0 \widehat{\nabla}^2 - \dd^2U_0/\dd y^2 \right) + \im \Rey^{-1} \widehat{\nabla}^4 \right] \hat{v} - \Fr^{-2} k_z \dd\hat{b}/\dd y &= \omega \widehat{\nabla}^2 \hat{v} , \label{leom1} \\
k_z \left(\dd U_0/\dd y\right) \hat{v} + \left( k_x U_0 + \im \Rey^{-1} \widehat{\nabla}^2 \right) \hat{\eta} - \Fr^{-2} k_x \hat{b}  &= \omega \hat{\eta} , \label{leom2} \\
- k^{-2} \left( k_z \dd\hat{v}/\dd y + k_x \hat{\eta} \right) + \left[ k_x U_0 + \im \left( \Rey\Sc \right)^{-1} \widehat{\nabla}^2 \right] \hat{b}  &= \omega \hat{b} , \label{leom3}
\end{align}
\end{subequations}
where $k^2=k_x^2+k_z^2$ is the wavevector squared norm and $\widehat{\nabla}^2=\dd^2/\dd y^2 - k^2$ is the Laplacian operator in Fourier space. System \rf{leom} describes a boundary eigenvalue problem of the form $A\hat{\bm{\xi}}=\omega B \hat{\bm{\xi}}$, for the eigenfrequency $\omega$ and eigenfunction $\hat{\bm{\xi}}=[\hat{v},\hat{\eta},\hat{b}]^{T}$ (the over-script denotes transposition). The latter set of equations is supplemented with no-slip and insulating boundary conditions, reading $\hat{v}=\dd\hat{v}/\dd y=\hat{\eta}=\dd\hat{b}/\dd y=0$ at $y=\pm 1$.

We use a Galerkin approach to discretize the differential operators, based on the expansion of the eigenfunctions in terms of Chebyshev polynomials. Simultaneously, we apply a collocation method at the Gauss--Lobatto quadrature nodes
\begin{equation}
\label{gln}
y_j = \cos{\left(\frac{j\pi}{M+1}\right)} , \quad j=1,\dots,M \:,
\end{equation}
for a fixed truncation order $M$. We verified the convergence of this method, based on the relative error of the eigenvalues of \rf{leom}, as $M$ was increased. In general, a value of $M\sim40$ was large enough to reach a reasonable tolerance.

\subsection{Results}
\label{Sec3b}

The classical unstratified ($\Fr\to+\infty$) plane Poiseuille flow is subject to a linear instability, due to the growth of a Tollmien--Schlichting (TS) wave \citep{T29,S33}. This well-known instability occurs at a critical Reynolds number $\Rey_c\sim5770$ \citep{O71} and {\color{black}{has been the subject of extensive studies on the transitionfrom laminar flows to turbulence}}. As shown in \citep{G21}, {\color{black}{the consideration of a stable profile of density stratification perpendicular to the shear plane  greatly lowers the instability threshold}}. The principal argument suggested by the authors is that such a flow allows for resonances of internal gravity waves (among themselves or eventually with viscous TS waves), leading to the onset of a new instability. A similar conclusion was drawn in the context of plane Couette flow, where a growth of perturbations due to the resonance of Doppler-shifted gravity waves was observed \citep{Couette18}. We recall though that Couette flow is unconditionally stable to infinitesimal disturbances in the unstratified limit and maintained by viscosity, which is not the case here.

\begin{figure}
\centering

\subfloat[$\Rey=550$]{
\includegraphics*[width=.32\textwidth]{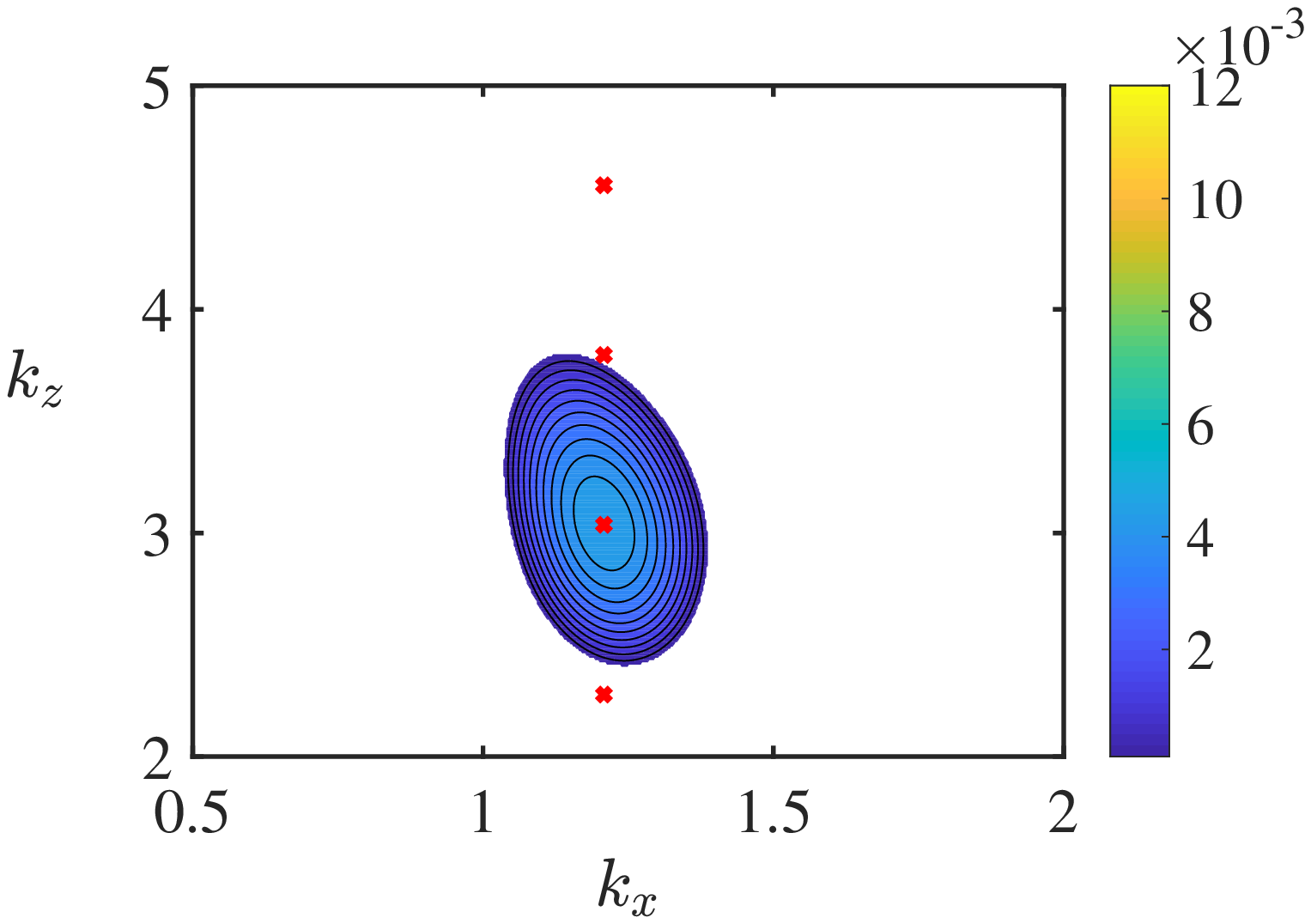}\label{Fig1a}} 
\subfloat[$\Rey=580$]{
\includegraphics*[width=.32\textwidth]{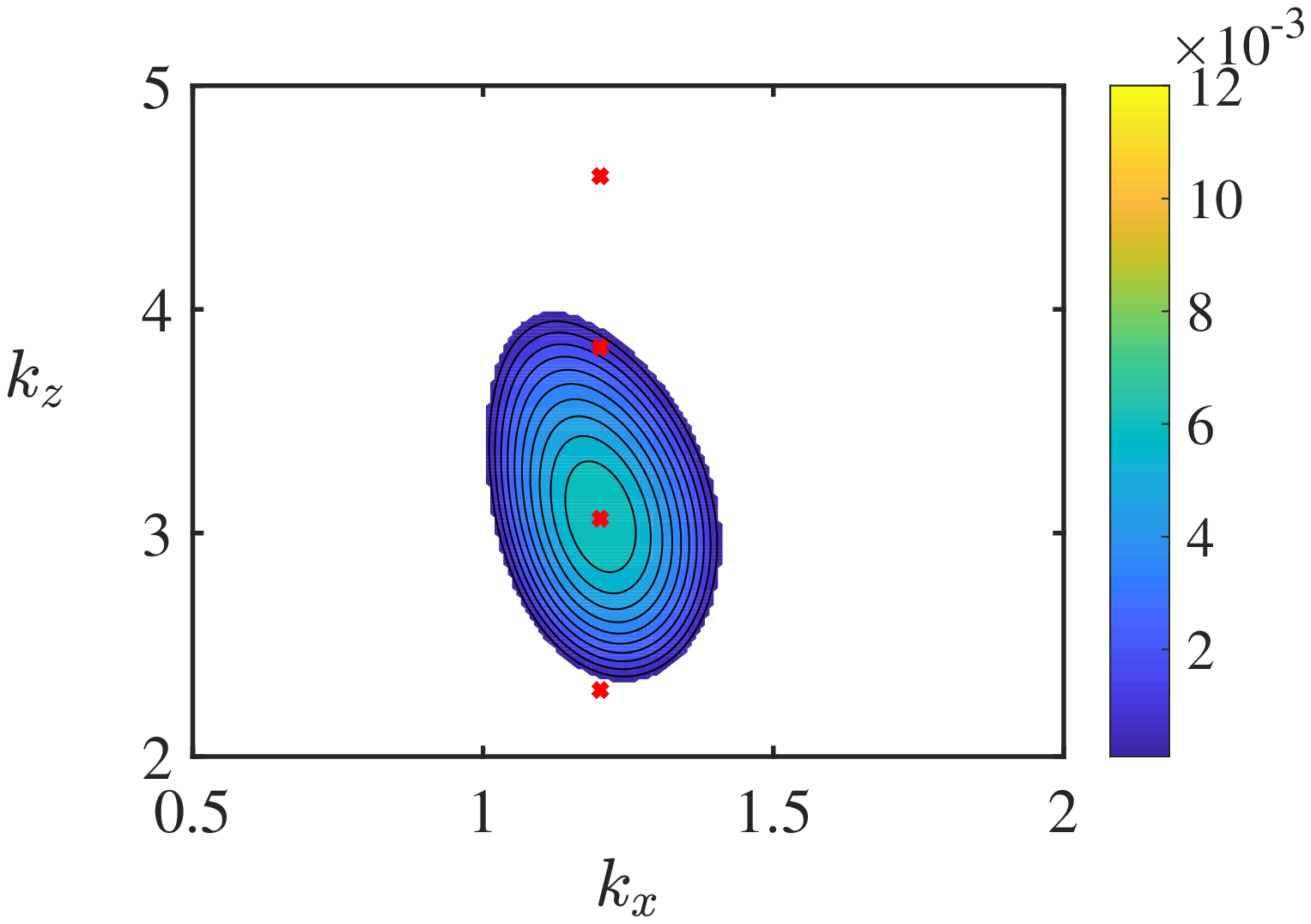}\label{Fig1b}} 
\subfloat[$\Rey=700$]{
\includegraphics*[width=.32\textwidth]{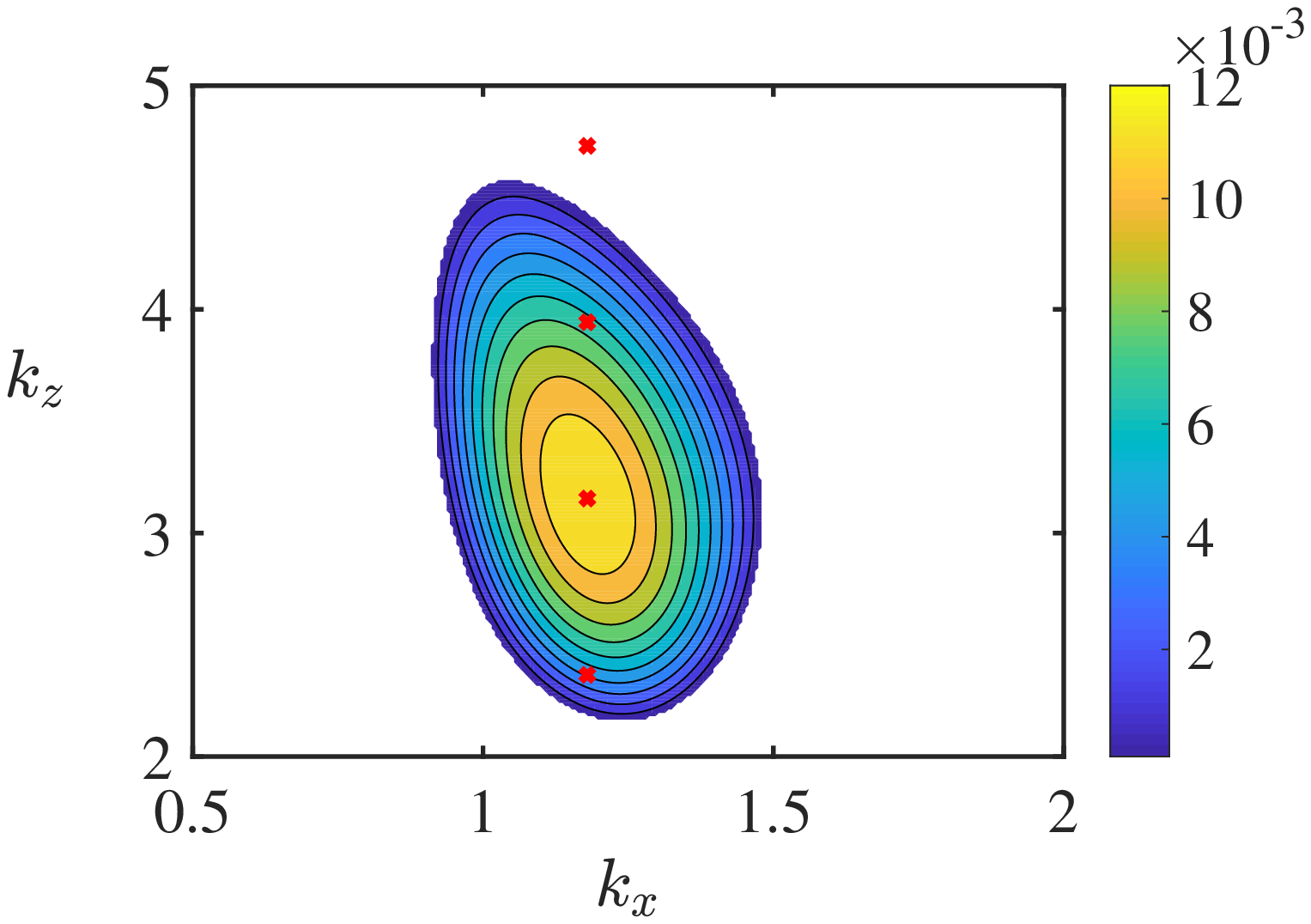}\label{Fig1c}}

\caption{Growth rate contours in the $(k_x,k_z)$-space from the linear stability analysis of equations \rf{leom} at $\Fr=2$, $\Sc=1$ and over different Reynolds numbers. Red crosses correspond to the discretized modes present in our DNS domain (cf. section \ref{Sec4}). White regions correspond to negative growth rates, i.e., stability.}
\label{stab_Re}
\end{figure}

Our interest for the present article is to go beyond the results obtained in \cite{G21}, by conducting a thorough analysis of the nonlinear saturation of this recently discovered instability. Therefore, we use the linear stability results to extract the optimal wavenumbers $(k_x^{\textrm{opt}},k_z^{\textrm{opt}})$, associated with modes of largest growth rates for a given set of control parameters. We display in figure \ref{stab_Re} the stability results at different Reynolds numbers, close to the instability threshold (for this set of parameters, we have a critical Reynolds number $Re_c\sim480$). We recall that $\Fr=2$ and $\Sc=1$ (differing thus from experimental values). For computational reasons, the red crosses in this figure represent the location in the wavenumber space of the discretized modes present in our DNS domain, see section \ref{Sec4} below. We then determine whether unstable harmonics of the fundamental mode are present within the unstable region or not. It becomes clear that the case $Re=550$ contains only stable harmonics, whereas the other two cases allow unstable harmonics within the domain of instability.

\section{Weakly nonlinear saturation near the onset} 
\label{Sec4}

{\color{black}{This section is devoted to the stability analysis of the nonlinear dynamical regime, in the vicinity of the bifurcation.}} As mentioned, the numerical domain chosen is an elongated rectangular parallelepiped of size $\mathcal{D} = [-\lambda_x/2,\lambda_x/2] \times [-1,1] \times [-2\lambda_z,2\lambda_z]$, with fundamental wavelengths $\lambda_x = 2\pi/k_x^{\textrm{opt}}$ and $\lambda_z = 2\pi/k_z^{\textrm{opt}}$ which depend on the control parameters. Our main focus being the possible emergence of modulations along the $z$-coordinate, this is why we consider boxes encompassing multiple unstable wavelengths along the vertical. However, we do not explore eventual horizontal modulations since we restrict the streamwise extension to only one unstable wavelength. We monitor the growth of the instability by computing the quantity
\begin{equation}
\label{KEz}
\mathcal{K}_z = \frac{1}{2} \langle w^2 \rangle_{\mathcal{D}} = \frac{1}{2\mathcal{D}} \int_{\mathcal{D}} w^2 \dd V ,
\end{equation}
where $\langle\cdot\rangle_{\mathcal{D}}$ denotes averaging over the volume $\mathcal{D}$. We therefore expect, from the definition of \rf{KEz} and linear theory, to first observe an exponential increase in energy, following a slope of twice the growth rate.

\begin{figure}
\centering

\subfloat[]{
\includegraphics*[width=.45\textwidth]{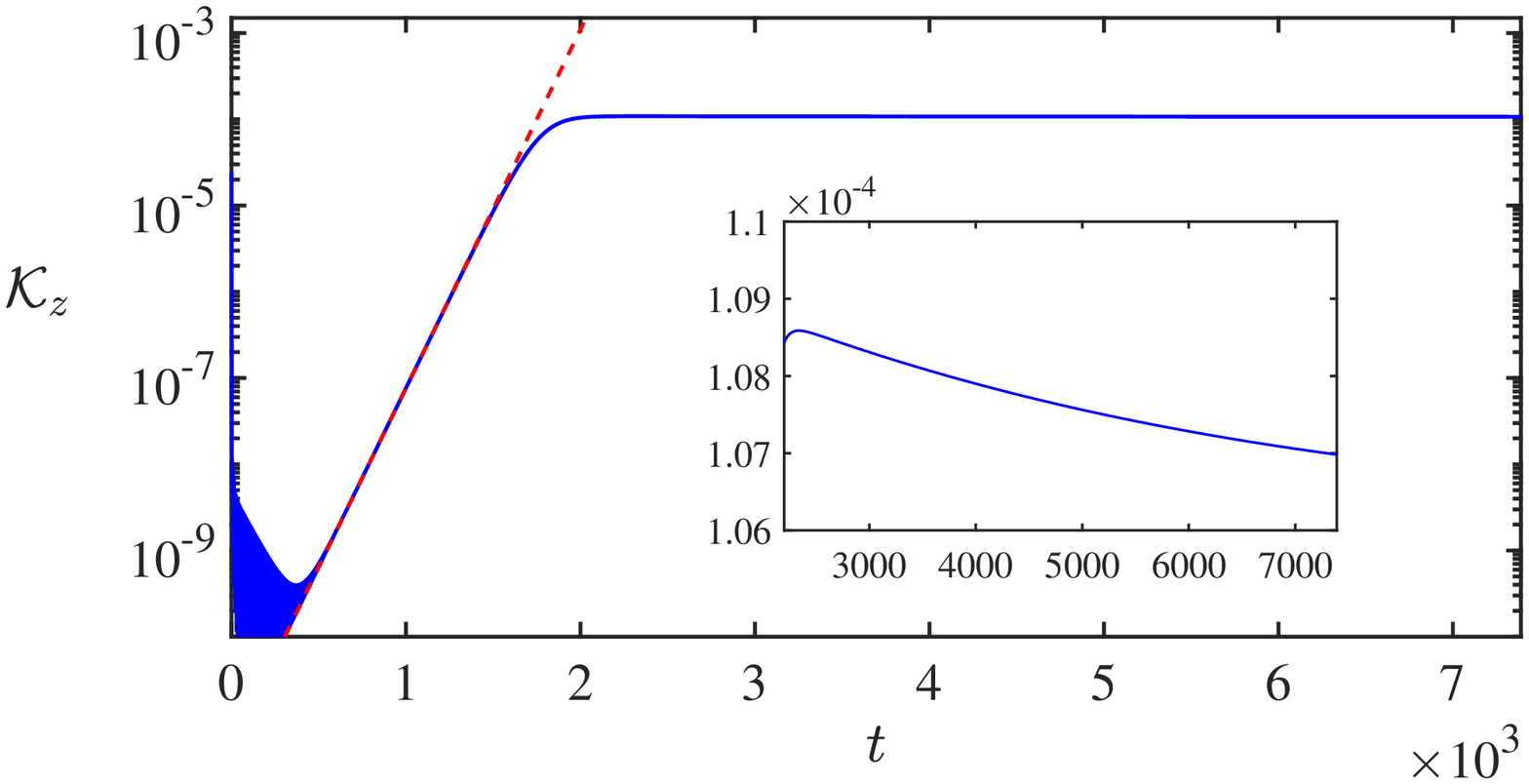} \hspace{1.2em}
\includegraphics*[width=.21\textwidth]{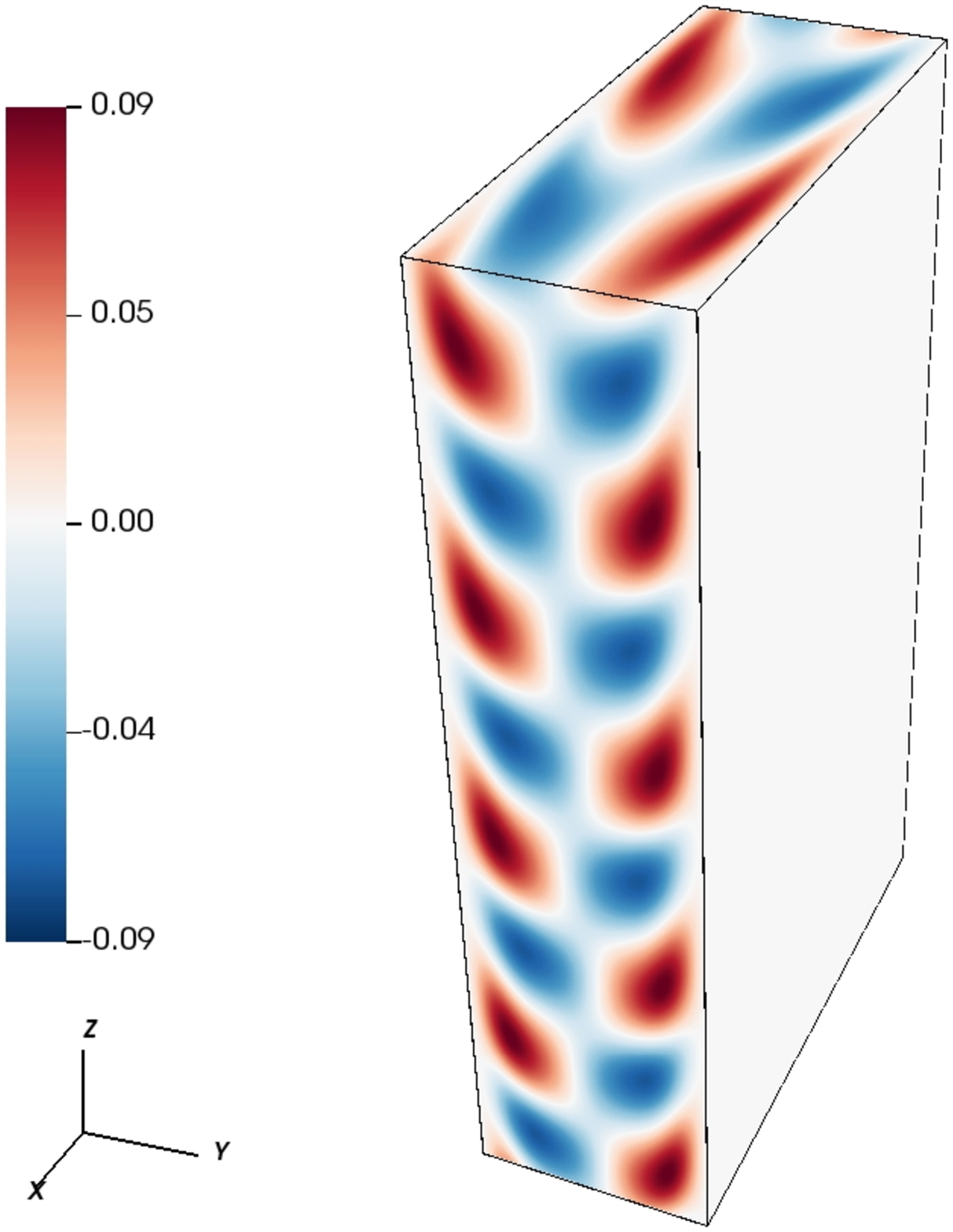} \hspace{1.2em}
\includegraphics*[width=.21\textwidth]{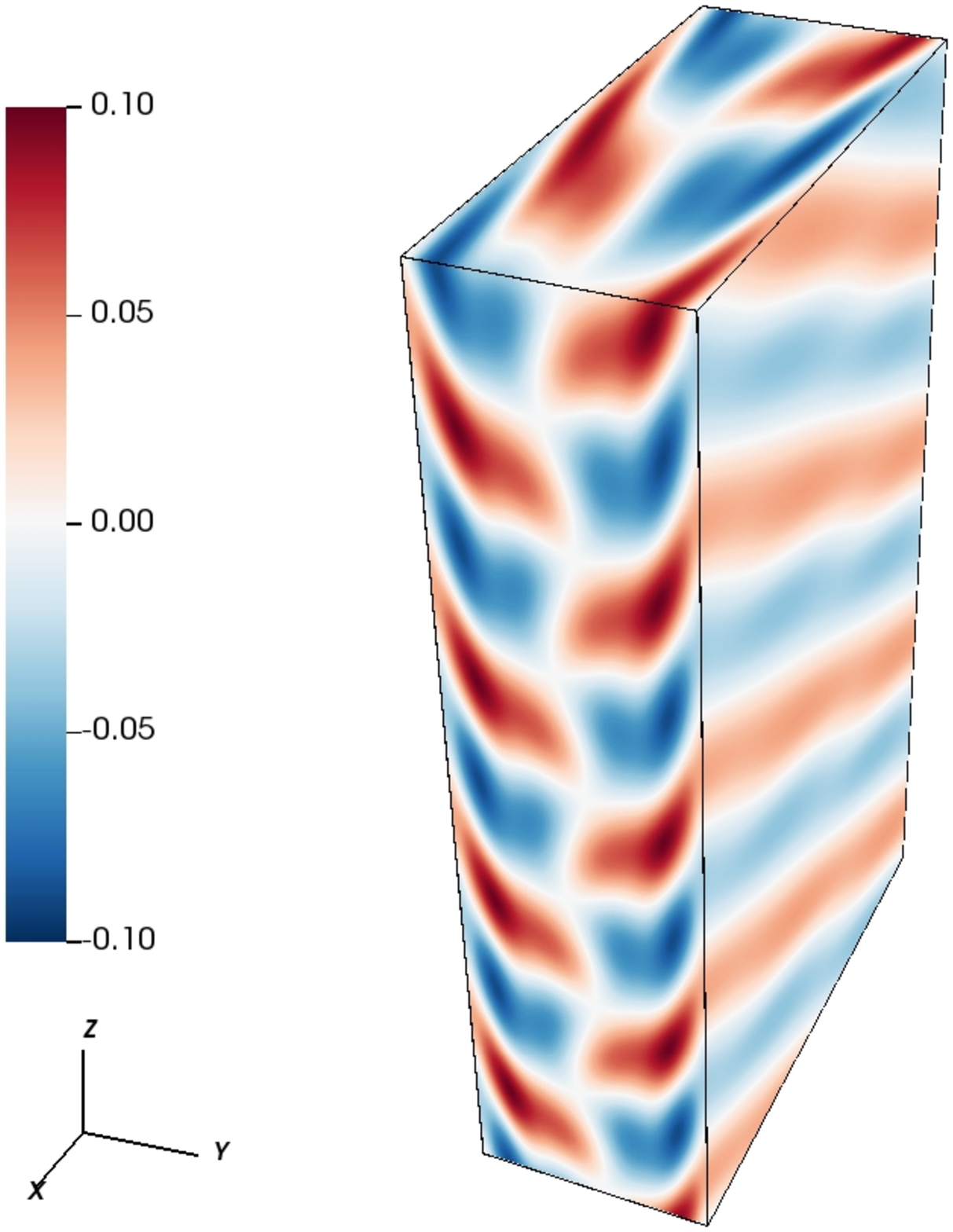}
\label{Fig2a}} \\
\subfloat[]{
\includegraphics*[width=.45\textwidth]{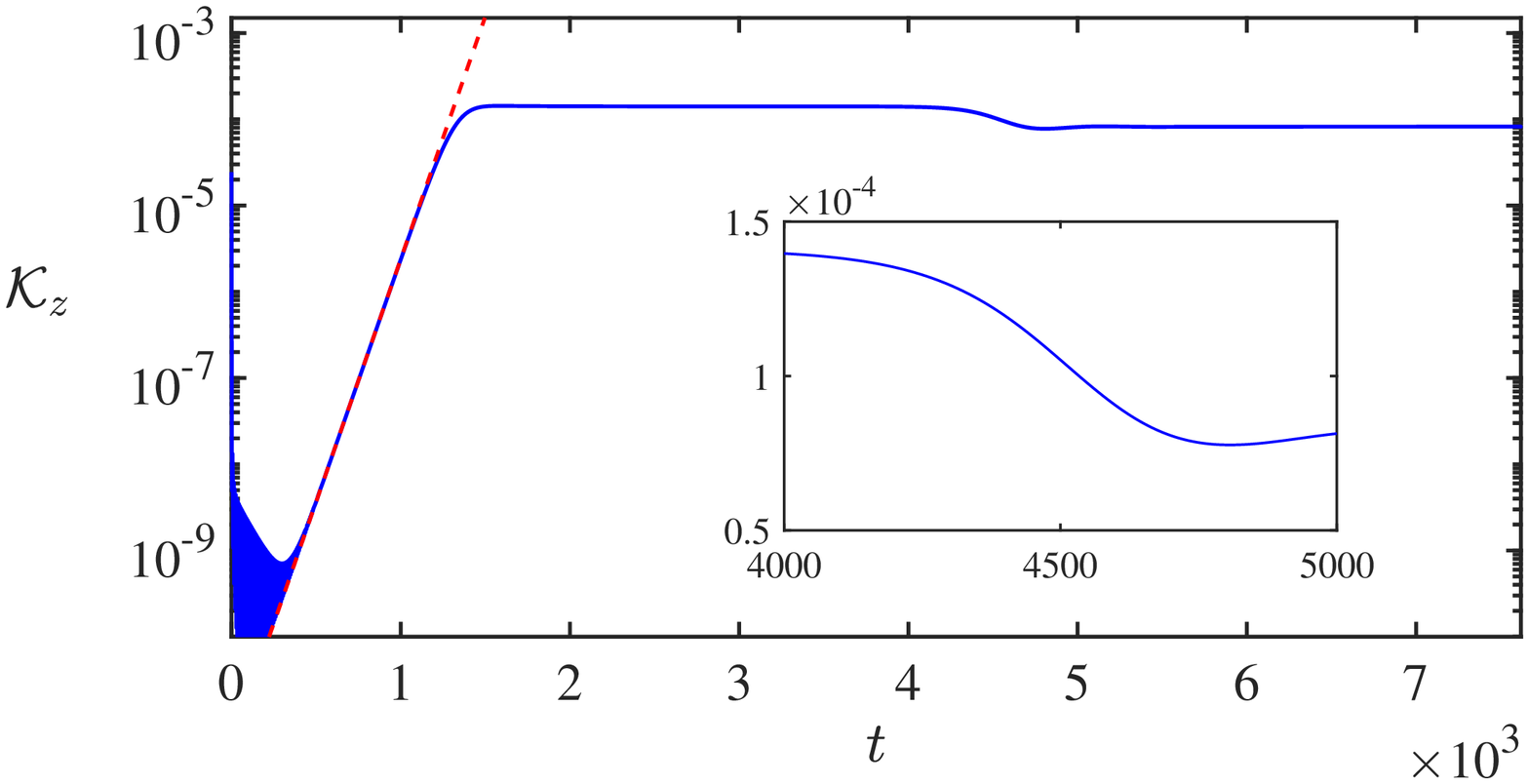} \hspace{1.2em}
\includegraphics*[width=.21\textwidth]{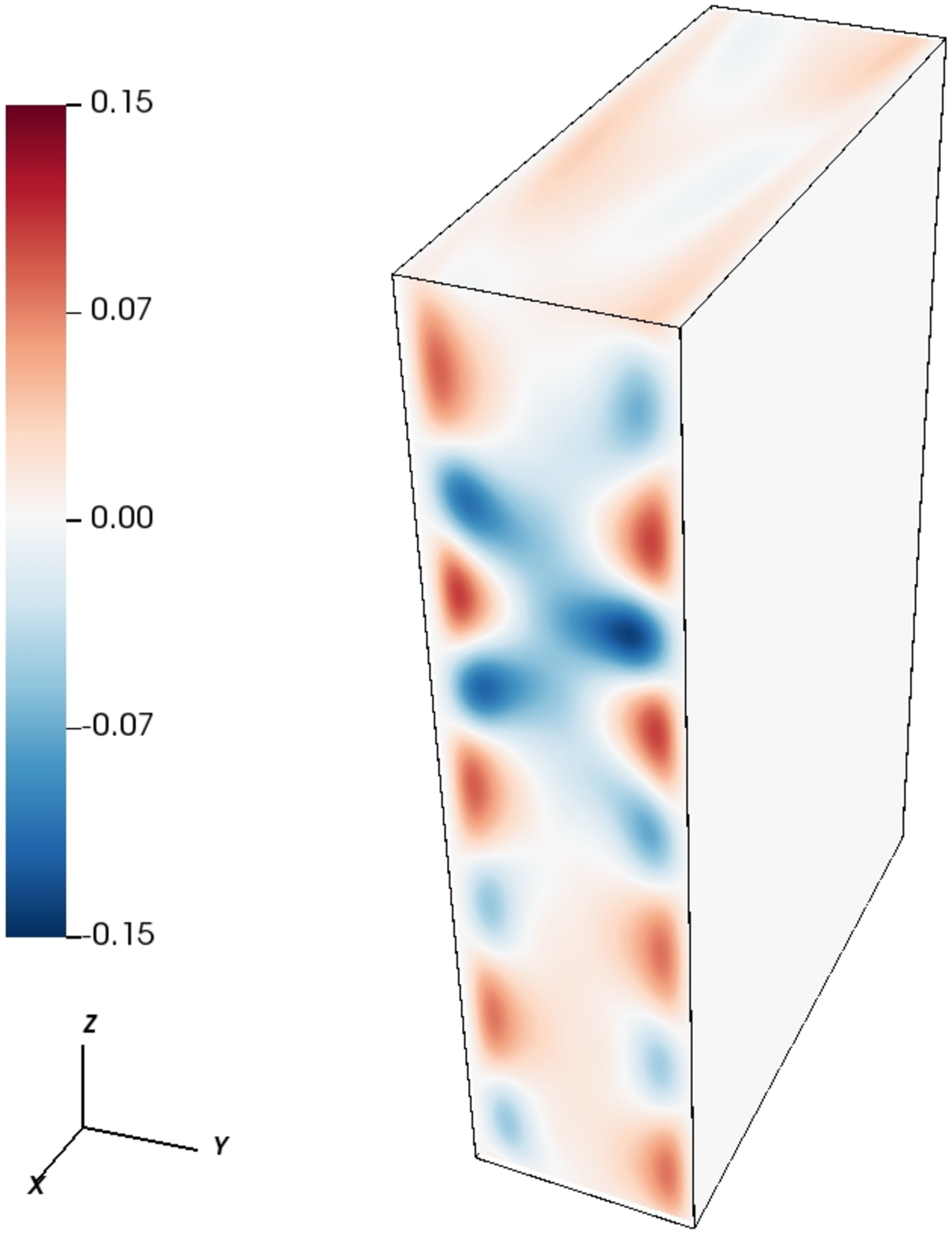} \hspace{1.2em}
\includegraphics*[width=.21\textwidth]{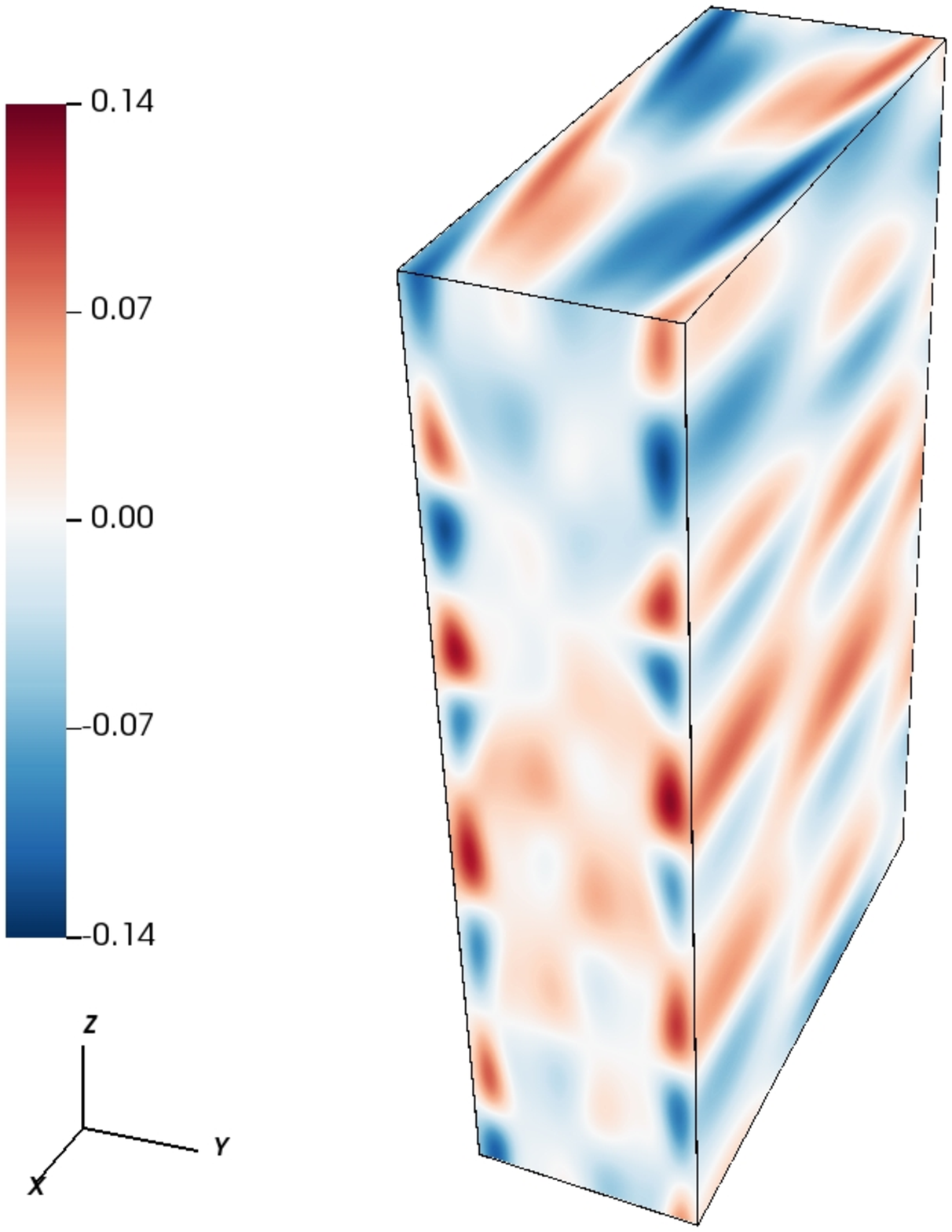}
\label{Fig2b}} \\
\subfloat[]{
\includegraphics*[width=.45\textwidth]{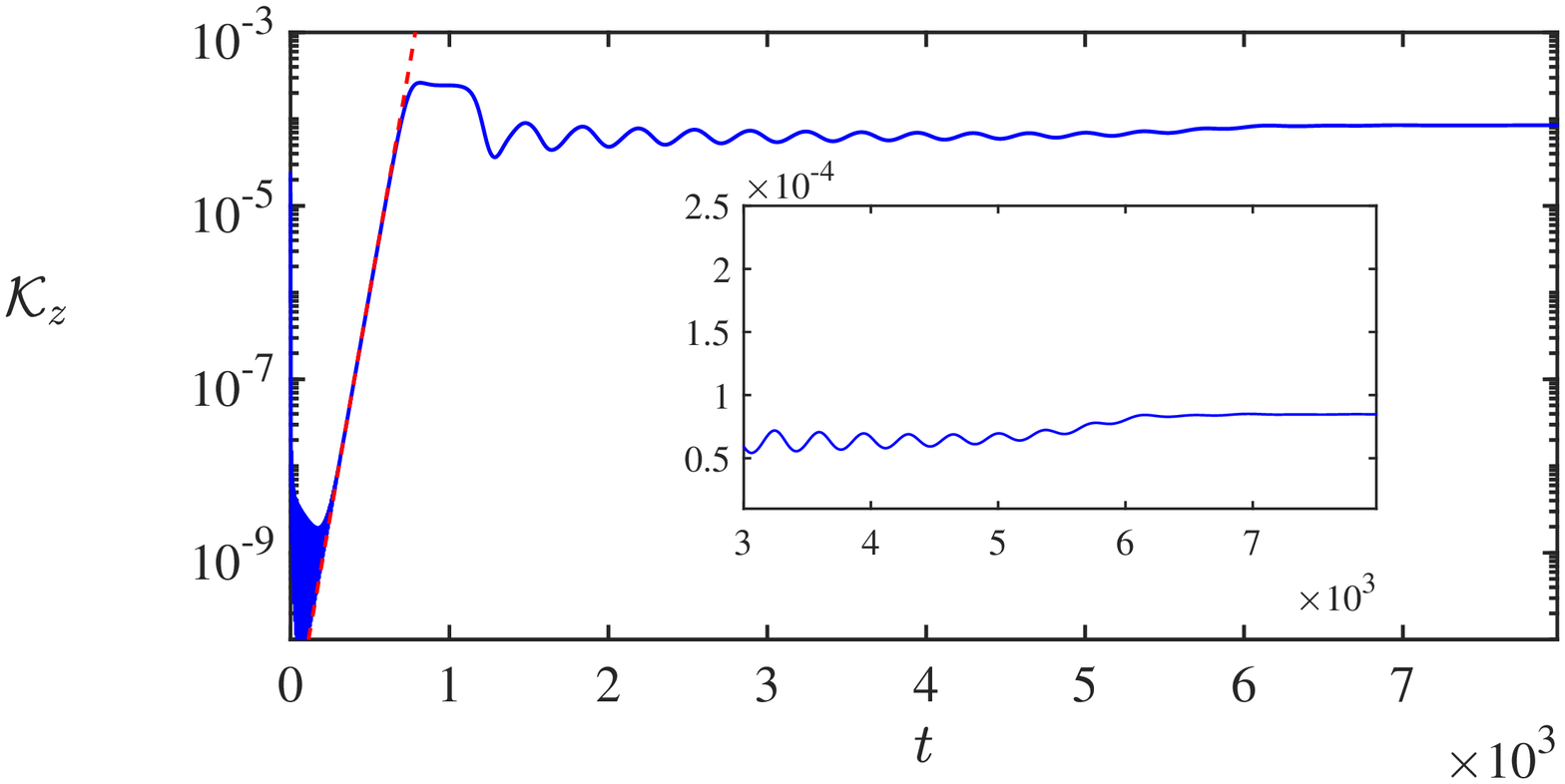} \hspace{1.2em}
\includegraphics*[width=.21\textwidth]{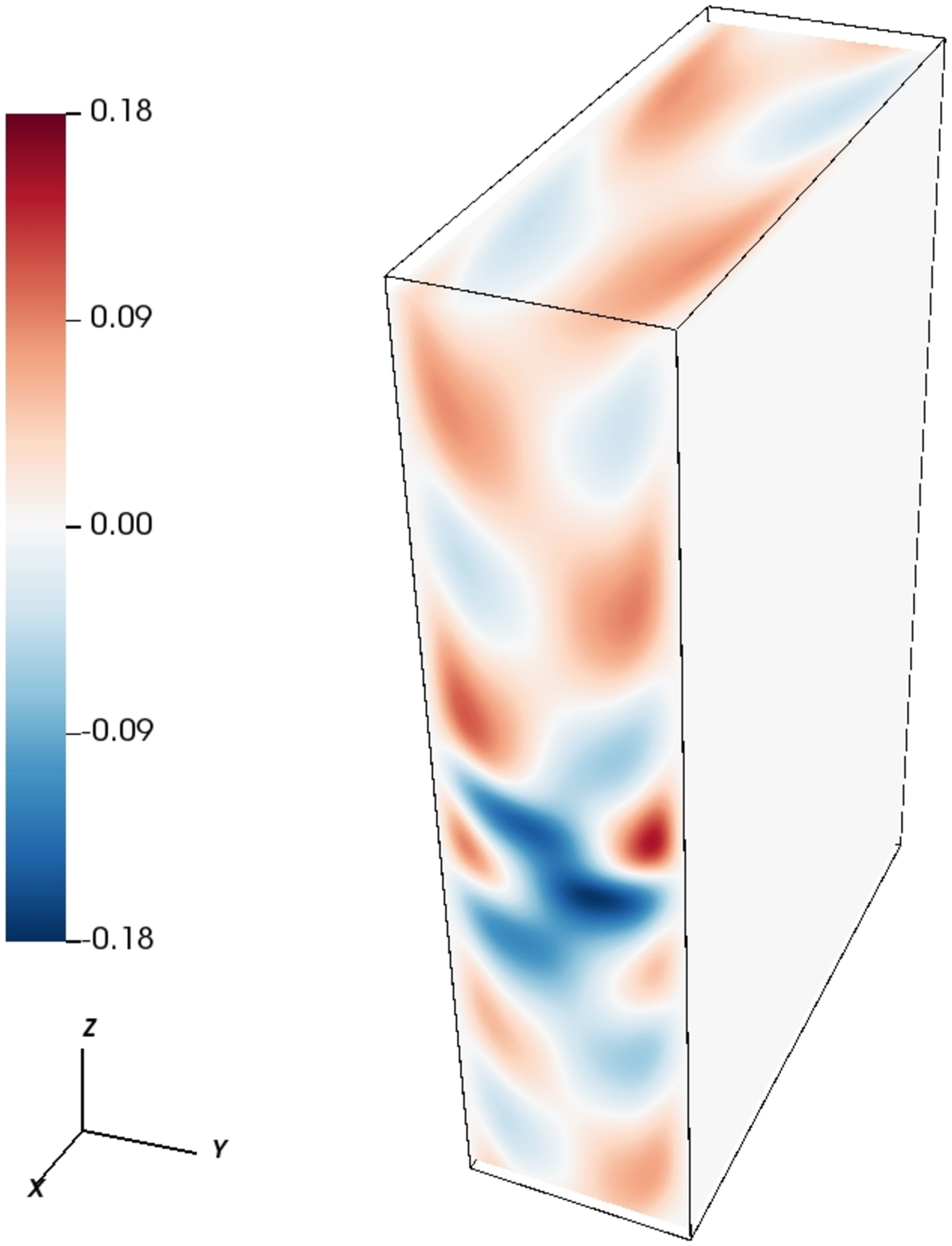} \hspace{1.2em}
\includegraphics*[width=.21\textwidth]{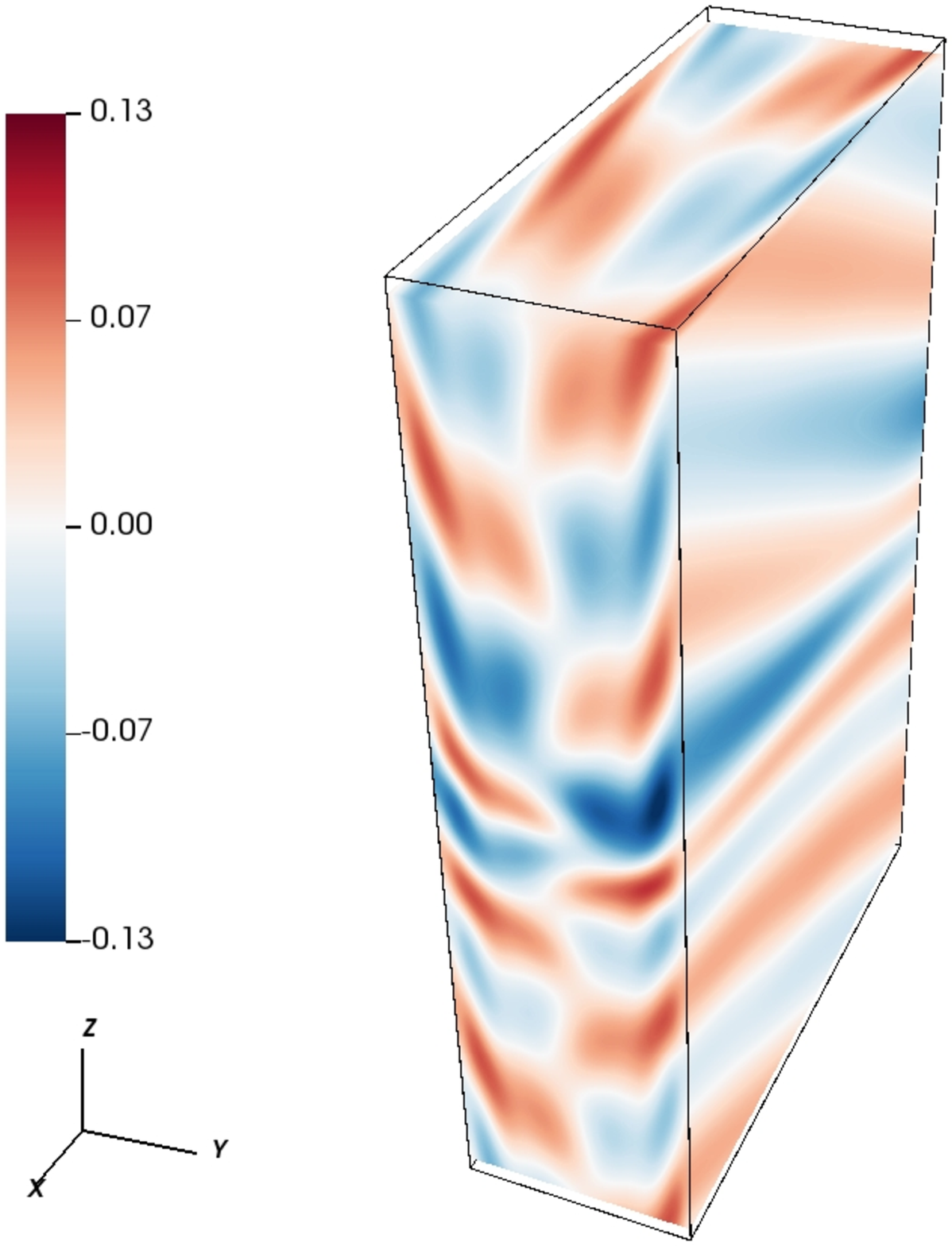}
\label{Fig2c}}

\caption{Vertical kinetic energy curves (in logarithmic scale) over time (left panels), with snapshots of the streamwise velocity fluctuation $u-U_0$ (middle panels) and the buoyancy perturbation $b$ (right panels), computed at the latest time of each simulation. Values of Reynolds numbers correspond to, respectively, (a) $\Rey=550$, (b) $\Rey=580$ and (c) $\Rey=700$. Red dashed lines in left panels represent twice the growth rates of the dominant unstable mode, as predicted from linear theory. Close-up views are displayed for each configuration, highlighting the nonlinear behaviour.}
\label{KZ1}
\end{figure}

We start by presenting in figure \ref{KZ1} a set of vertical kinetic energies, computed over time, for Reynolds numbers close to the instability threshold (in the present case, $Re_c\sim480$). In addition, we add snapshots of the streamwise velocity and perturbed buoyancy profile at the latest instant of computation (last point of the energy curves). Not surprisingly, we notice the exponential growth of disturbances in the first part of each curve. The slope in {\color{black}{logarithmic}} scale is further confirmed from linear stability analysis, as shown by the red dashed lines. Then, once the perturbations reach an order of magnitude similar to that of the base flow, nonlinearities are no longer negligible and the system departs from linear theory. 

The first case of interest, $Re=550$, is fairly simple to understand as the instability saturates in the form of a quasi-stationary solution (we still nevertheless observe a small decay over a viscous time in the nonlinear regime as shown by the insert in figure \ref{Fig2a}). The saturated amplitude represented here corresponds physically to a meandering in the streamwise velocity, according to the velocity snapshot displayed in the same figure. {\color{black}{Similar profiles have already been observed in the experiment of \cite{G21} (cf. figure 3a in this reference for PIV and DNS visualizations)}}. Besides this steady and vertically homogeneous configuration, we notice interesting features in the other two panels, corresponding to the computations at $Re=580$ and $Re=700$. There is indeed a finite time at which this homogeneous state departs from the primary branch via a secondary bifurcation, as displayed in the inserts of figure \ref{KZ1}. Once the bifurcated solutions reach a stationary state, at the end of each simulations, the corresponding velocity and buoyancy profiles are no longer homogeneous. As a consequence, a localized structure of velocity emerges along the vertical and the buoyancy perturbations also become inhomogeneous and localized. Clearly, we observe a spontaneous pattern formation due to a symmetry-breaking mechanism, our system being initially invariant along $z$.

\begin{figure}
\centering

\subfloat[$\Rey=550$]{
\includegraphics*[width=.9\textwidth]{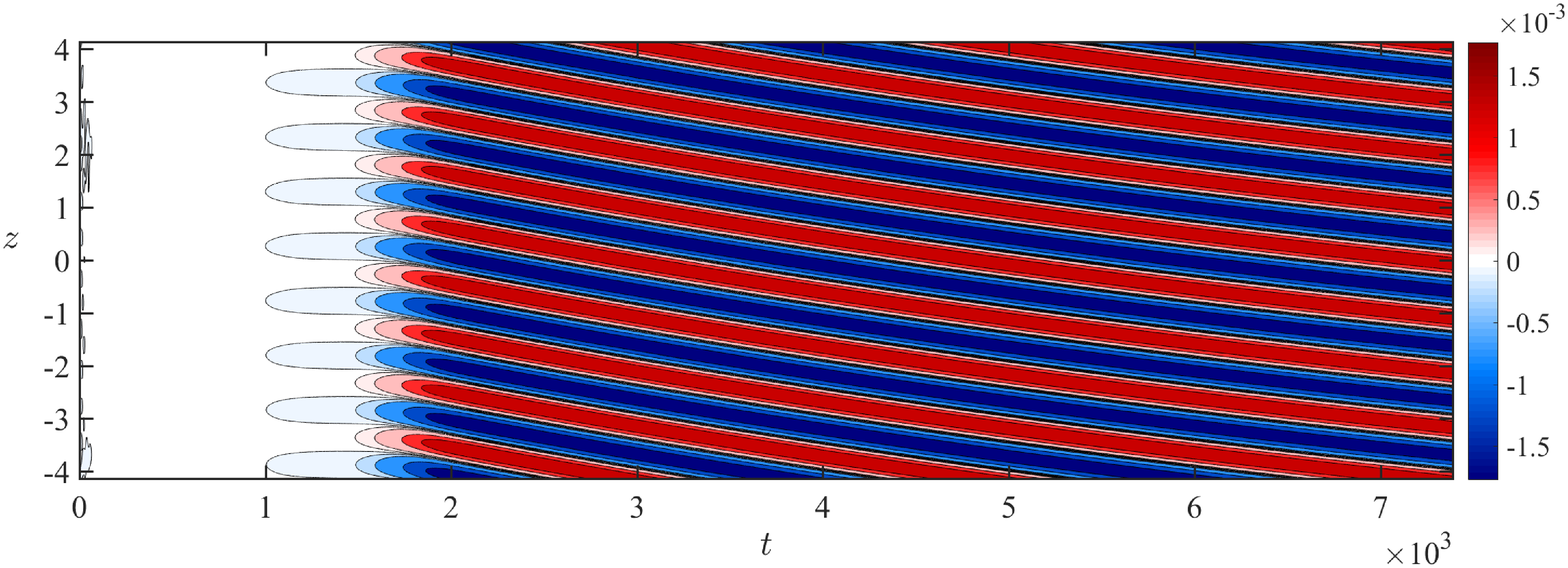}\label{Fig3a}} \\ \vspace*{-1.2em}
\subfloat[$\Rey=580$]{
\includegraphics*[width=.9\textwidth]{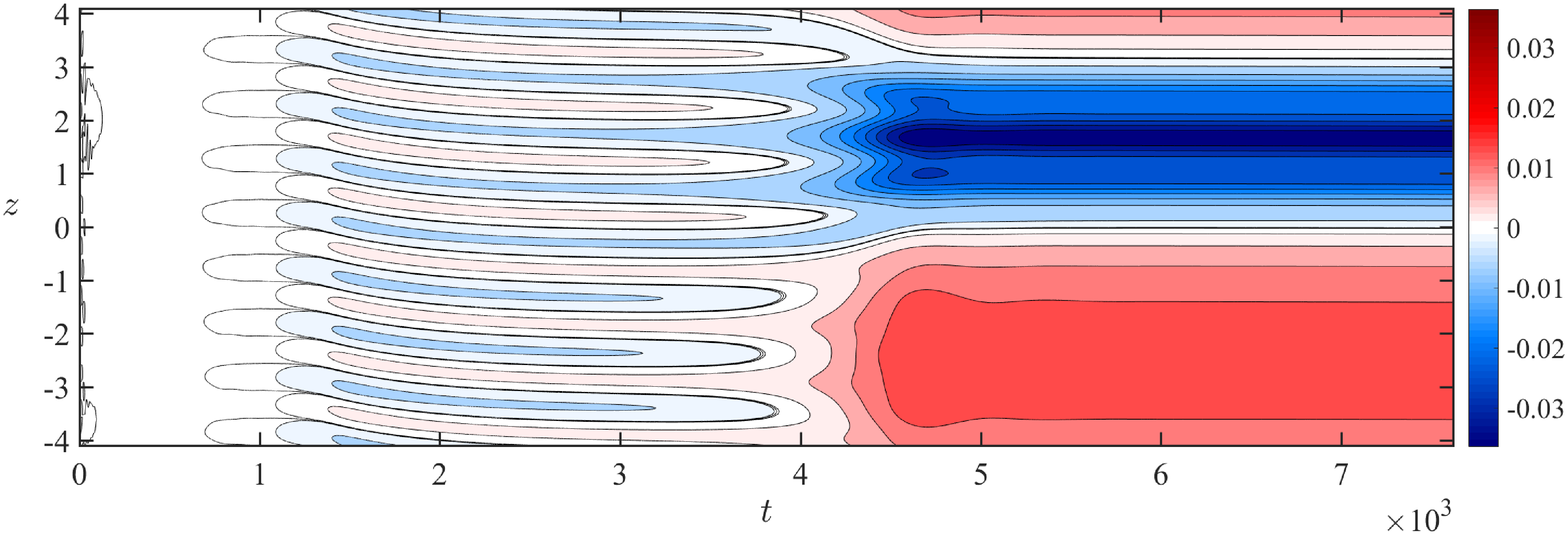}\label{Fig3b}} \\ \vspace*{-1.2em}
\subfloat[$\Rey=700$]{
\includegraphics*[width=.9\textwidth]{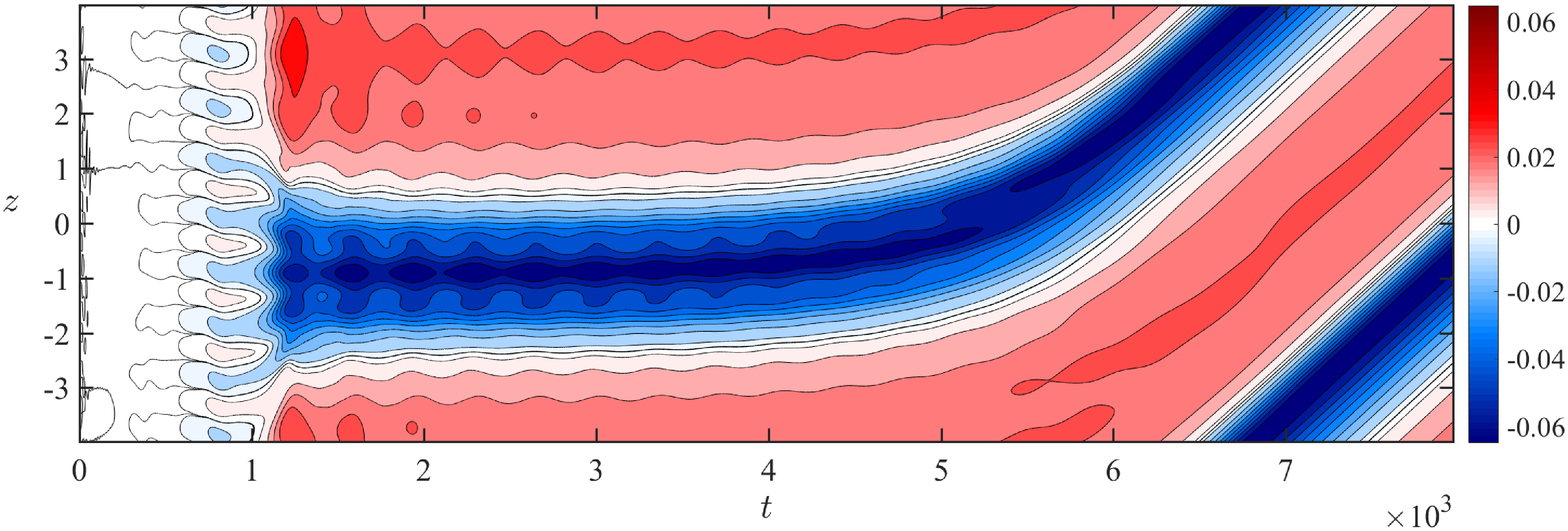}\label{Fig3c}}

\caption{Hovm\"oller diagrams of the horizontally averaged streamwise velocity fluctuation $\langle u - U_0 \rangle_H$, computed in $(t,z)$-plane. Positive (negative) values are displayed in red (blue). The areas of constant color on the left of each panels represent the exponential growth of disturbances.}
\label{HD1}
\end{figure}

{\color{black}{As we notice in most of the cases a loss of symmetry along the vertical, we introduce an averaging operator for quantifying horizontal measures}}, defined as $\langle\cdot\rangle_H \coloneqq (2\lambda_x)^{-1} \iint\cdot\,\dd x\dd y$. One way to capture and identify the peculiar transition described earlier is to compute Hovm\"oller diagrams of the horizontally-averaged streamwise velocity fluctuation $\langle u - U_0 \rangle_H$. This is represented in figure \ref{HD1} for the three configurations we are interested in. As we already noticed earlier, the case corresponding to $\Rey=550$ displays harmonic modulations of the horizontally-averaged streamwise velocity which is slowly drifting in time. Note that this is effectively a correction to the initially $z$-invariant Poiseuille profile, although it remains small in amplitude at this Reynolds number. For this simulation, we expect the weak temporal drift to vanish as we reach a large enough viscous time (scaling thus with $\Rey$), until the symmetric solution would ultimately be restored. This neutral drift observed here essentially depends on the initial random seed used for the computation. Indeed, we observed various shifts of the solution depending on the initial perturbations we used (not shown). Pursuing our investigation, the most striking observation from figure \ref{Fig3b} is the sudden coarsening of the mean flow modulation, at the finite time corresponding to the secondary bifurcation in figure \ref{Fig2b}. The system indeed undergoes for $t\sim4\times10^3$ a symmetry-breaking mechanism, allowing for the formation of a box-scale modulation of the mean Poiseuille profile that is not harmonic: the region of reduced streamwise velocity is sharp and localized while the accelerated region is broader (see figure \ref{ux_Re} below). We recall that the streamwise mass flux is conserved in all of our simulations. This phenomenon fills the whole numerical domain and saturates in the form of a steady nonlinear solution. More surprisingly, the case of $Re=700$ displayed in figure \ref{Fig3c} also depicts the same symmetry-breaking mechanism (although at a shorter time), with a localized structure in the velocity profile, but converges instead to a drifting mode. This uniform shift seems {\color{black}{unrelated}} with the drift observed at $\Rey=550$, since we observe it irrespective of the initial conditions. {\color{black}{Determining its origin would require a nonlinear analysis, which is beyond the scope of this paper}}. We recall here that a very similar phenomenology is observed when using the more classical imposed pressure gradient forcing to sustain the mean Poiseuille flow. It is naturally emerging from the nonlinear interaction of unstable waves with the mean flow, regardless of the force sustaining it.

\begin{figure}
\centering
\subfloat[]{
\includegraphics*[width=.465\textwidth]{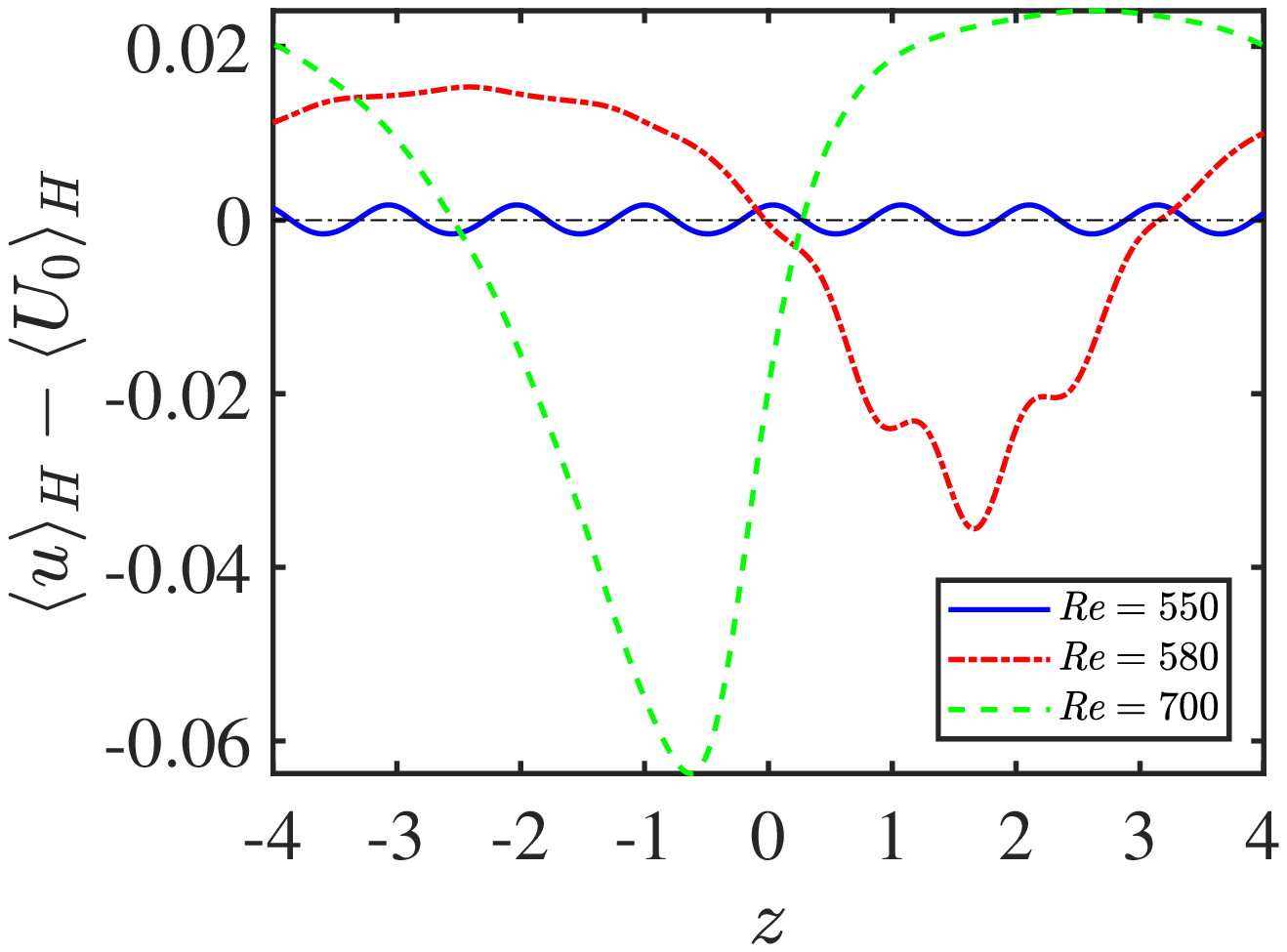}
\label{Fig4a}}
\subfloat[]{
\includegraphics*[width=.49\textwidth]{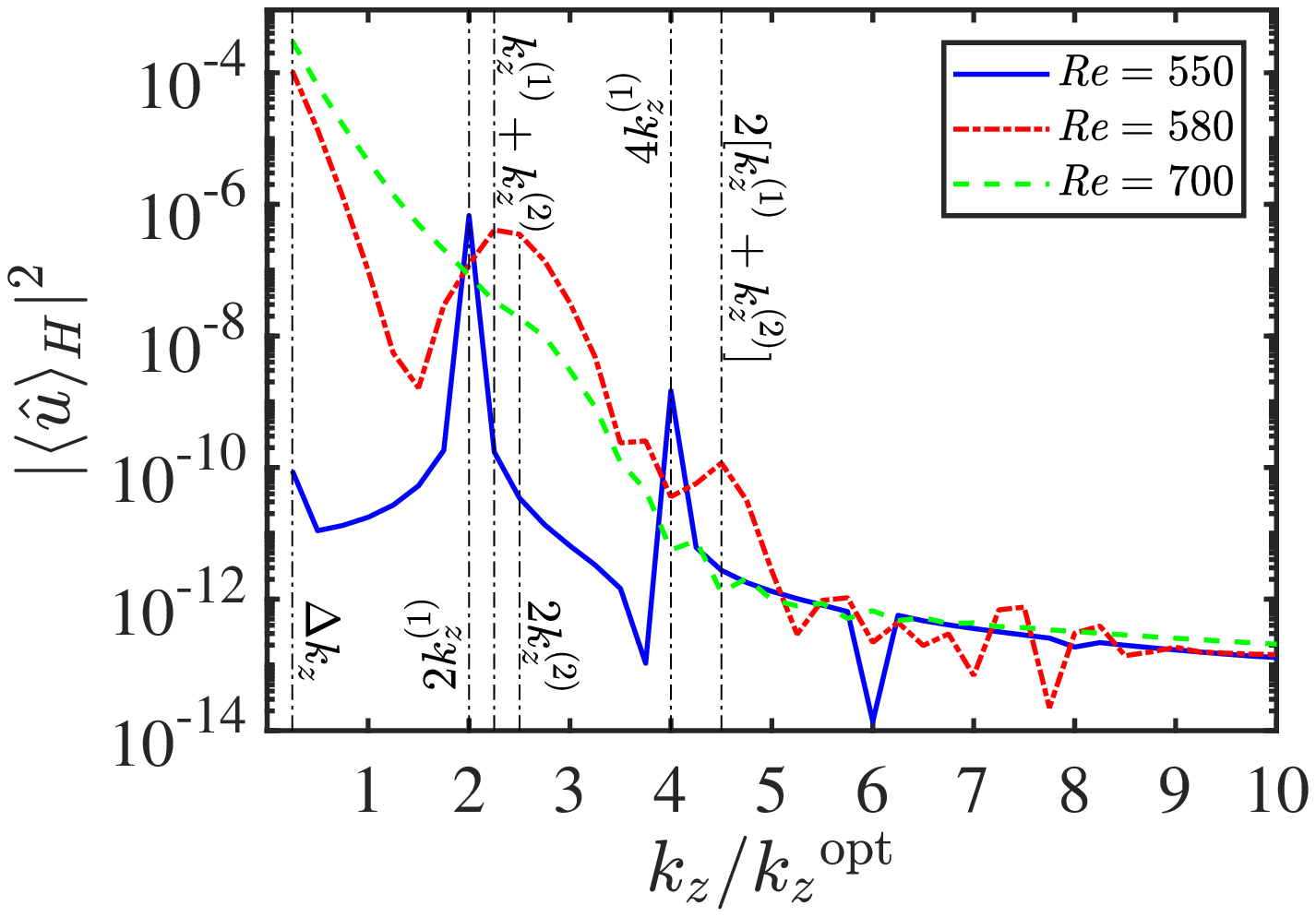}
\label{Fig4b}}
\caption{Each curve is computed at the latest time of simulation in figure \ref{HD1}, once the solution has converged. (a) Horizontally-averaged streamwise velocity fluctuations as a function of the vertical coordinate. (b) Spectral energy in Fourier space of the averaged velocity profiles from the left panel. The principal amplitude peaks are displayed with vertical lines and captioned with their associated representation defined in expression \ref{reystr}.}
\label{ux_Re}
\end{figure}

Let us now investigate the wave-mean flow interaction that is depicted here, {\color{black}{using}} arguments from linear theory. Using the averaging operator over the horizontal plane introduced earlier, we define the Reynolds-averaged decomposition of the field $\bm{q}=(\bm{u},p,b)$, as
\begin{equation}
\label{qrans}
\bm{q}=\langle\bm{q}\rangle_H(z,t) + \bm{q}'(x,y,z,t) ,
\end{equation}
separating thus the horizontal mean flow from the fluctuations. Substituting \rf{qrans} in the streamwise projection of equation \rf{eom1}, along with boundary conditions, we recover the mean field equation
\begin{equation}
\label{meaneq}
\frac{\partial}{\partial t} \langle u \rangle_H + \frac{\partial}{\partial z} \langle u' w' \rangle_H = \Rey^{-1} \left( \frac{\partial^2}{\partial z^2} \langle u \rangle_H + \langle \frac{\partial^2u}{\partial y^2} \rangle_H \right) + \langle f \rangle_H .
\end{equation}
Equation \rf{meaneq} expresses the evolution of the mean flow and the transfer of momentum, by means of the divergence of Reynolds stresses (the second term in the left-hand side). If we consider the fluctuations to be written as a superposition of both upward and downward modes with distinct but {\color{black}{nearby}} wavenumbers $(k_z^{(1)},k_z^{(2)})$ and frequencies $(\omega^{(1)},\omega^{(2)})$, we obtain
\begin{align}
\label{udec}
u' = \hat{u}_{11} &e^{\im(k_x x + k_z^{(1)} z - \omega^{(1)} t)} + \textrm{c.c.} + \hat{u}_{12} e^{\im(k_x x - k_z^{(1)} z - \omega^{(1)} t)} + \textrm{c.c.} \nn \\
&+ \hat{u}_{21} e^{\im(k_x x + k_z^{(2)} z - \omega^{(2)} t)} + \textrm{c.c.} + \hat{u}_{22} e^{\im(k_x x - k_z^{(2)} z - \omega^{(2)} t)} + \textrm{c.c.} ,
\end{align}
with the same decomposition holding for $w'$. Subsequently, we expand the Reynolds stress term in \rf{meaneq}, using the modal decomposition \rf{udec}, and we find
\begin{align}
\langle u' w' \rangle_H = &\langle \hat{u}_{11}\hat{w}_{11}^{*} + \hat{u}_{12}\hat{w}_{12}^{*} + \hat{u}_{21}\hat{w}_{21}^{*} + \hat{u}_{22}\hat{w}_{22}^{*} \rangle_H + \textrm{c.c.} \nn \\
&+ \langle \hat{u}_{11}\hat{w}_{12}^{*} + \hat{u}_{12}^{*}\hat{w}_{11} \rangle_H e^{2\im k_z^{(1)} z} + \textrm{c.c.} + \langle \hat{u}_{21}\hat{w}_{22}^{*} + \hat{u}_{22}^{*}\hat{w}_{21} \rangle_H e^{2\im k_z^{(2)} z} + \textrm{c.c.} \nn \\
&+ \langle \hat{u}_{11}\hat{w}_{22}^{*} + \hat{u}_{22}^{*}\hat{w}_{11} \rangle_H e^{\im[(k_z^{(1)}+k_z^{(2)}) z - \Delta\omega t]} + \textrm{c.c.} \nn \\
&+ \langle \hat{u}_{11}\hat{w}_{21}^{*} + \hat{u}_{21}^{*}\hat{w}_{11} \rangle_H e^{\im(\Delta k_z z - \Delta\omega t)} + \textrm{c.c.} \label{reystr},
\end{align}
{\color{black}{where $^{*}$ denotes}} complex transposition and for $\Delta k_z = k_{z}^{(1)} - k_{z}^{(2)}$ and $\Delta\omega = \omega^{(1)} - \omega^{(2)}$  (a similar description with only one pair of modes can be found in \cite{Y22}). We distinguish in \rf{reystr} the contributions from self- and cross-interactions between a fixed set of discretized modes. As one can notice, computing the vertical divergence of the whole expression removes the contribution of self-interacting modes (the first term in the right-hand side) from the mean flow equation. However, cross-interaction of modes with the same vertical structure (second and third terms of the expression) yields a wave pattern with twice the original wavenumber. {\color{black}{Expression \rf{reystr} allow us to demonstrate, in a discrete framework, the resonance mechanism and the modulation in the mean flow due to the Reynolds stresses.}}

We represent in figure \ref{ux_Re} snapshots of the averaged velocity fluctuations $\langle u - U_0 \rangle_H$ (we recall that $U_0=1-y^2$ is the plane Poiseuille profile), at the latest time of computations for each panel in figure \ref{HD1}. Clearly, we notice the loss of homogeneity, originally present in the solution at $\Rey=550$, when observing the other two curves. For instance, the transient solution $\Rey=580$ still displays the wave pattern at twice the initial wavenumber from the later case, but also tends to form a modulation that fills the entire numerical box. Additionally, we explore the spectral energy of the solution at, first, $\Rey=550$ in figure \ref{Fig4b} and notice, as expected, the two peaks at $2k_z^{\textrm{opt}}$ and its harmonics (with the correspondence $k_z^{\textrm{opt}}=k_z^{(1)}$ from expression \rf{reystr}). We can easily notice this pattern to emerge in the non-linear regime of the first simulation at $\Rey=550$, as displayed in figures \ref{Fig3a} and \ref{Fig4a}, with a total wavenumber of $8k_z^{\textrm{opt}}$ (since the domain originally contains 4 unstable wavelengths). This same spatial structure is also present in the early stages of figures \ref{Fig3b} and \ref{Fig3c}, until the time where the other modes start to contribute to the dynamics and break the symmetry. The intermediate case, $\Rey=580$, shows a transient solution with a leading mode of wavenumber $\Delta k_z = k_z^{\textrm{opt}}/4$ present in the spectrum of figure \ref{Fig4b}. Moreover, there exists lower peaks corresponding to $k_z^{(1)}+k_z^{(2)}=9k_z^{\textrm{opt}}/4$, $2k_z^{(2)}=5k_z^{\textrm{opt}}/2$, with their harmonics, but they all are dominated by the mode at the size of the whole domain. Finally, the spectrum of the latest configuration at $\Rey=700$ is completely dominated by the mode of wavenumber $\Delta k_z$ and enclosed by two sharp fronts of velocity. This profile is represented in figure \ref{Fig4a} and highlights the generation and amplification of vertical shear, due to the increasing velocity gradient in that direction (with an order of magnitude of $\sim1$ to $10\%$). We conclude that this spontaneous large-scale modulation is thus mainly due to an interaction of closely spaced unstable harmonics, generating interesting space-time wave patterns. Not surprisingly, these symmetry-breaking secondary bifurcations happen at a shorter time as the control parameter ($Re$) increases, allowing indeed more harmonics within the instability map. A demonstration of this feature was represented using linear theory in figure 1, with red crosses depicting the different harmonics. 

{\color{black}{This modulation that leads to a localized flow pattern could be compared to the exact invariant solutions of the Navier--Stokes equations as discovered in shear flows without stratification \citep{GibsonBrand}. However, today it is not known if the solutions that we discover in our numerical simulations have any connection with these localized nonlinear solution of the unstratified Poiseuille flow. To answer this open question, a continuation approach should be performed in order to follow the bifurcated non linear branches when changing the Froude and Reynolds numbers. This task, which shares some interesting connections with the case of the rotating plane Couette flow \citep{Nagata}, is of course beyond the scope of the present study.}}

We now try exploring configurations away from the instability threshold, to investigate whether we still observe a similar phenomenology and how it transits to turbulence.

\section{Localized stratified turbulence far from the onset}
\label{Sec6}

\begin{figure}
\centering

\subfloat[]{
\includegraphics*[width=.8\textwidth]{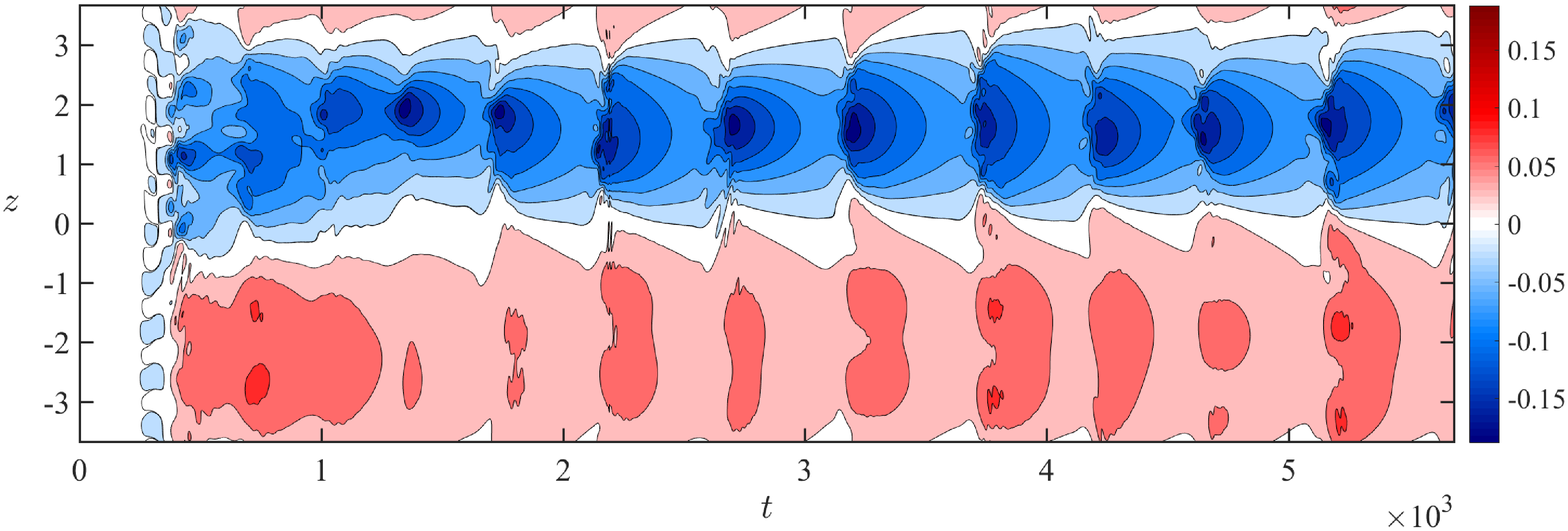}\label{Fig5a}} \\ \vspace*{-1em}
\subfloat[]{
\includegraphics*[width=.8\textwidth]{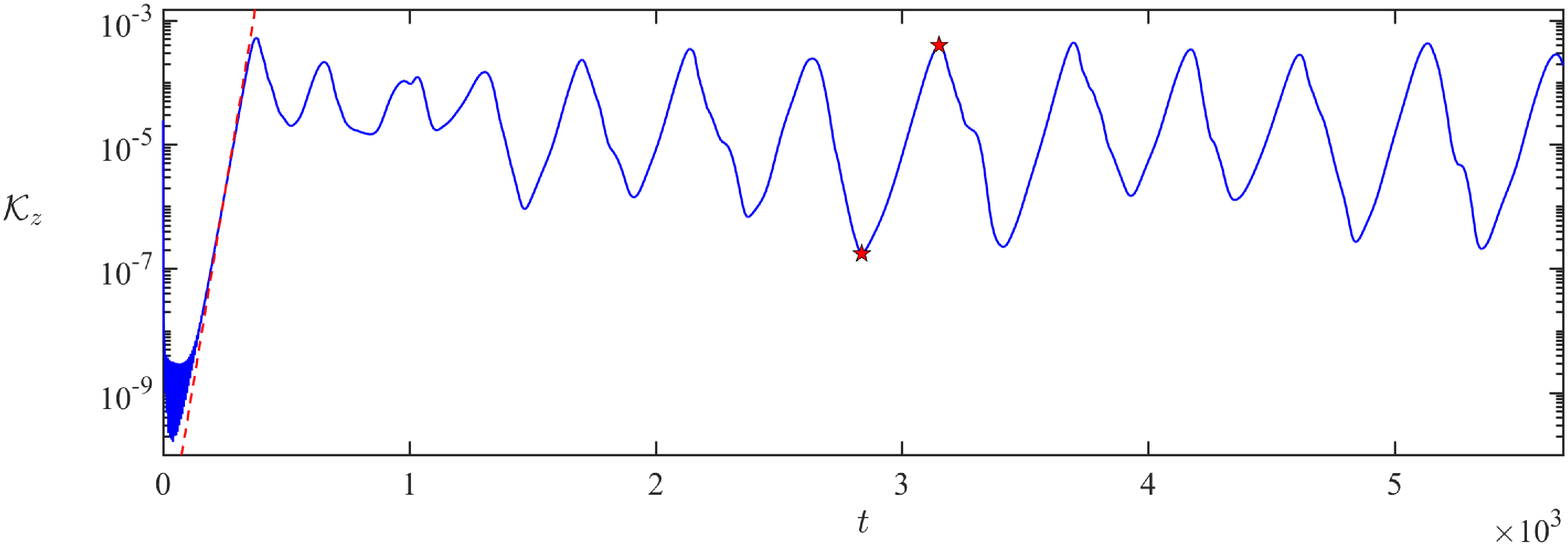}\label{Fig5b}} \\ \vspace*{-1em}
\subfloat[]{
\includegraphics*[width=.8\textwidth]{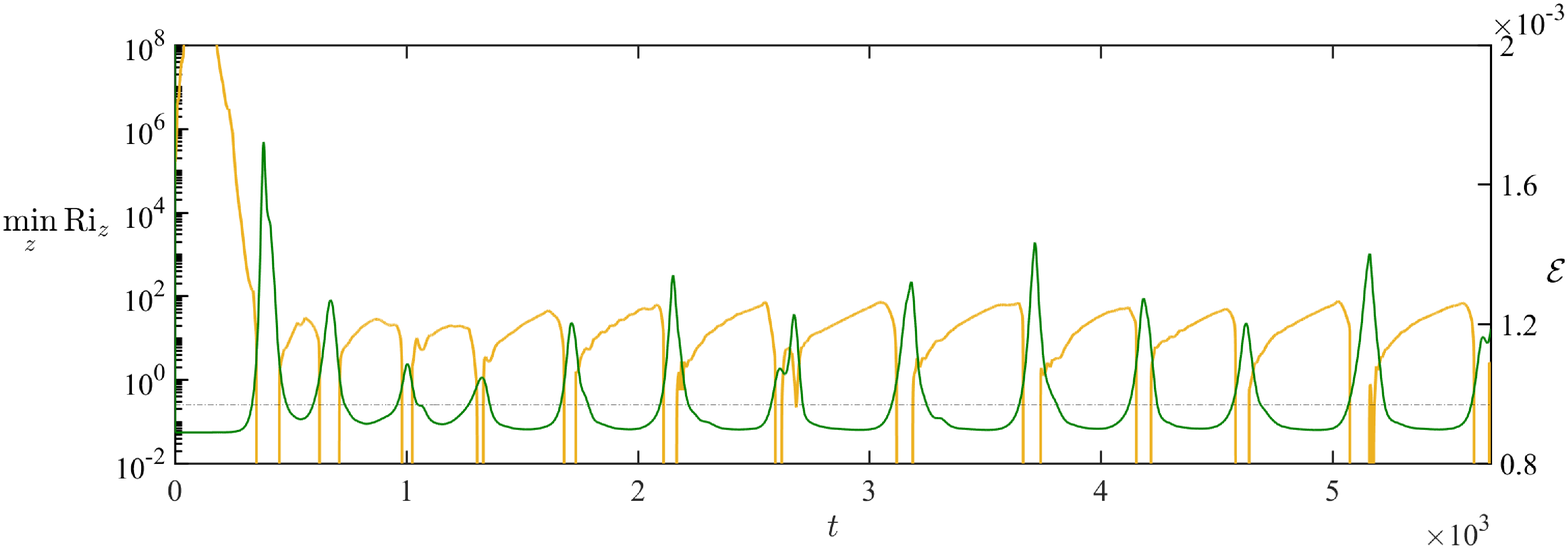}\label{Fig5c}}

\caption{DNS at $Re=1500$. (a) Hovm\"oller map of streamwise mean velocity fluctuations, with the same color code as in figure \ref{HD1}. (b) Vertical kinetic energy in the spirit of figure \ref{KZ1}. The two red markers indicate the extrema of a bursting episode, computed in details in figure \ref{R1500burst}. (c) Vertical minimum of gradient Richardson number over time, computed in logarithmic scale (left $y$-axis, in yellow) and viscous dissipation $\mathcal{E}$ (right $y$-axis, in dark green). The red curve is discontinuous whenever $\Ri_z$ is negative (when the flow becomes locally unstable to convection). The thin dashed-dotted line delimits the KH instability threshold of $\Ri_z=1/4$ \citep{M61,H61}.}
\label{R1500map}
\end{figure}

Departing away from the instability threshold, we expect the flow to behave in a more chaotic way and we notably seek for the trigger of stratified turbulence. As already noticed in a previous study on the plane Couette profile with spanwise stratification \citep{LCK19}, {\color{black}{the onset of turbulence occurs when $Re\gg Re_c$ in regions with largest vertical velocity gradients (we recall that $Re_c$ is a critical Reynolds number computed from linear theory)}}. It appears that the mechanism responsible for these events {\color{black}{is induced by a secondary Kelvin--Helmholtz (KH) type overturning instability}} \citep{M61,H61,LCK19}. We therefore pay great attention in our computations on the local quantities determining the tendency of the flow to be subject to secondary shear instabilities. In particular, we establish local expressions for the Froude and gradient Richardson numbers \citep{LCK17}, as follows
\begin{equation}
\label{FrRiz}
\Fr_z = \frac{\Fr}{\sqrt{1 - \Fr^2\langle \partial_z b\rangle_H}}, \quad \Ri_z = \frac{1}{\Fr_z^2\langle \partial_z u\rangle_H^2} .
\end{equation}
It is worth emphasizing that our system is initially homogeneous in the vertical direction, such that $\left.\Fr_z\right|_{t=0}=\Fr$ and $\left.\Ri_z\right|_{t=0}=\infty$. {\color{black}{In addition to these quantities, we also compute the viscous dissipation 
\begin{equation}
\label{diss}
\mathcal{E} = \Rey^{-1} \left< \vert\bm{\nabla}\times\bm{u}\vert^2 \right>_{\mathcal{D}} ,
\end{equation}
where the quantity within brackets is the total enstrophy of the flow.}}

We represent the local Richardson number in figure \ref{R1500map}, along with the vertical kinetic energy and the Hovm\"oller diagram for the case $Re=1500$, highlighting the saturation of the linear Poiseuille instability and the appearance of a non-trivial dynamics. First, we still observe the large-scale vertical modulation of the mean Poiseuille profile discussed in the previous section. This phenomenon is therefore robust and persists well beyond the weakly nonlinear regime. Second, we observe in all the panels a series of bursting episodes, consisting in cycles of {\color{black}{energy growth and decay}}. Soon after the symmetry-breaking transition at $t\sim300$, the local Richardson number in figure \ref{Fig5c} drastically falls until reaching a small interval of time where it becomes negative. This peculiarity emphasizes the presence of unstable stratification locally and supports the onset of a secondary shear instability. We study further the quasi-periodic patterns displayed over time in figures \ref{Fig5b} and \ref{Fig5c}, and explain this series of events as a consequence of the loss of vertical homogeneity, leading to the growth of vertical shear. Indeed, once the Richardson number becomes low enough to cross the theoretical threshold of $\Ri_z=1/4$ \citep{M61,H61}, the flow strengthens due to this secondary instability, until saturation, and then collapses into unsteady motion. {\color{black}{We use the classical linear threshold as an indication only since we consider an unsteady nonlinear regime. Note however that the Richardson number drops over several order of magnitude in a very short time.}} Afterwards, energy diminishes due turbulent dissipation {\color{black}{(displayed in green)}} and presumably because the secondary instability has disappeared due to a mixing event, or the reduction in vertical shear. {\color{black}{Indeed, the correlation between the fall in Richardson number and the growth of viscous dissipation soon after indicates that a secondary vertical shear-induced instability triggers first and is then followed by a turbulent collapse and increased dissipation. The mean flow is then partially restored (still modulated along the vertical) to a state dominated by the interaction of internal gravity waves. In practice, the global picture might be more complex than this interpretation, due to the superposition of the two instabilities and the three-dimensional character of the flow, but all the previous arguments tend to support our overall description of this mechanism.}} From a dynamical point of view, we can view the system as trapped in a quasi-periodic attractor. Similar observations were made in the context of forced stratified turbulence, in relation with the formation of density layers \citep{LCK17}.

\begin{figure}
\centering

\subfloat[]{
\includegraphics*[width=.98\textwidth]{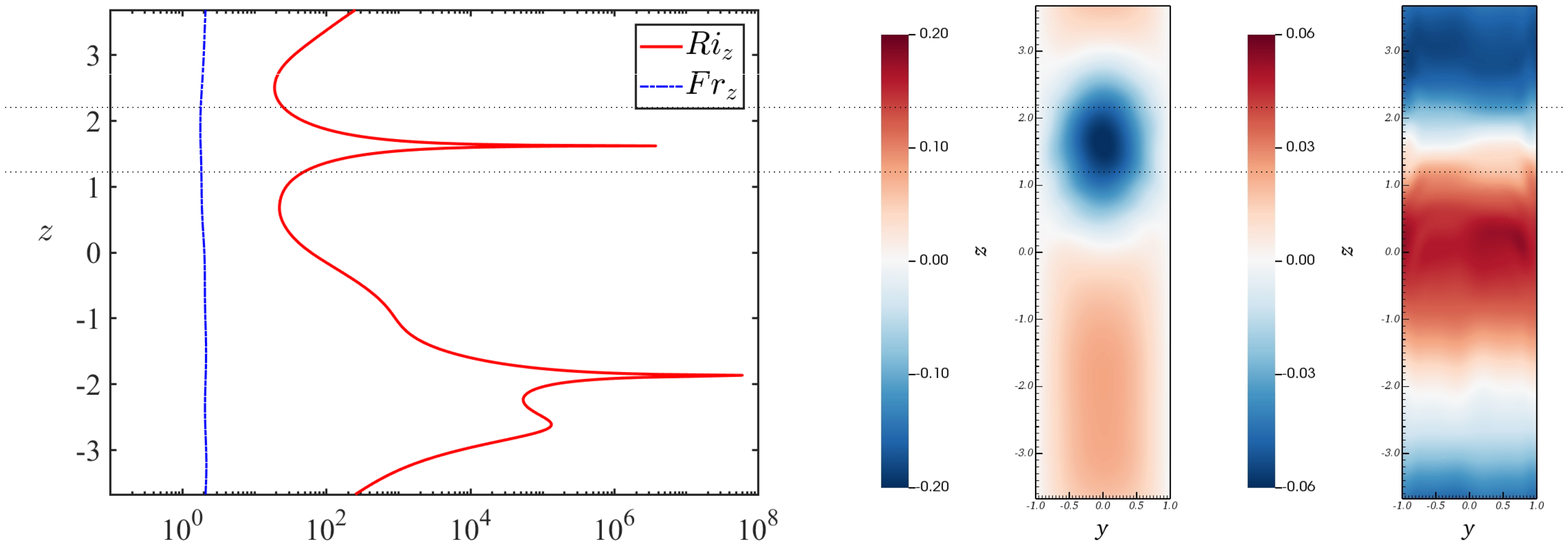}
\label{Fig6a}
} \\
\subfloat[]{
\includegraphics*[width=.98\textwidth]{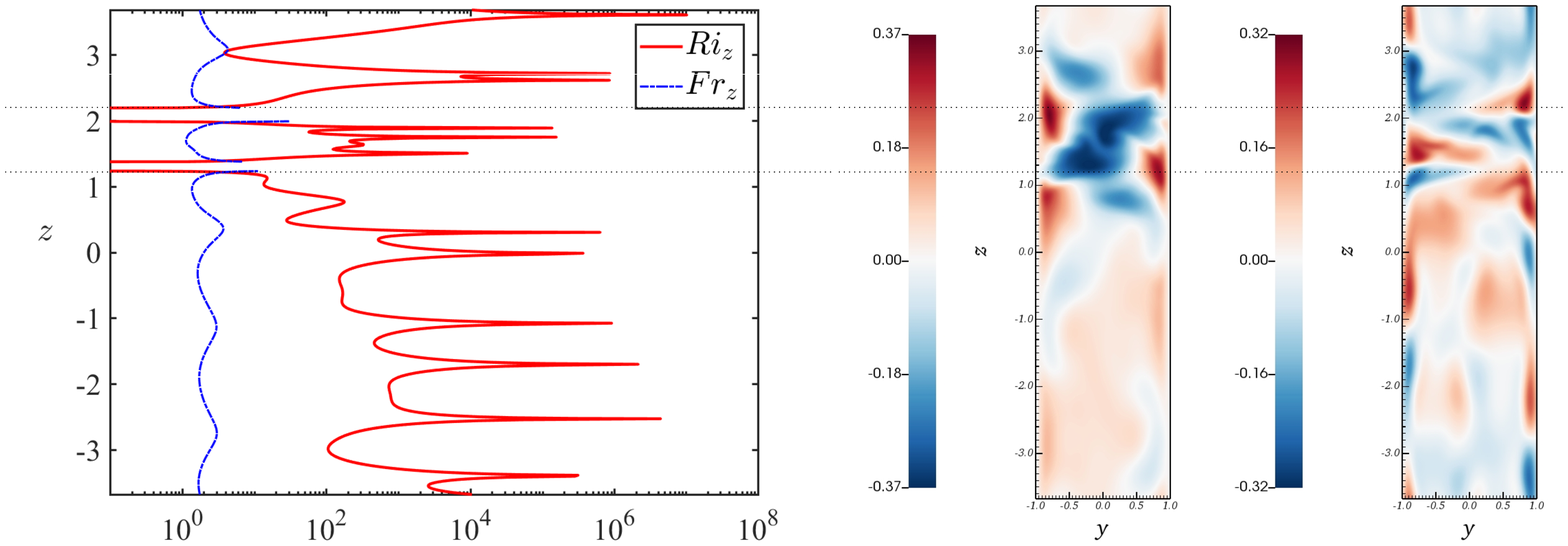}
\label{Fig6b}
}

\caption{Local measures of Froude and Richardson numbers (in logarithmic units) over the vertical coordinate, for the case $\Rey=1500$. Data and snapshots are computed at (a) $t=2837$ and  (b) $t=3151$ (see figure \ref{R1500map}). Second and third panels of each rows represent the streamwise velocity fluctuation $u-U_0$ and the perturbation of buoyancy $b$, respectively, both displayed at an arbitrary position along $x$. The upright straight line in the left panel indicates the threshold of KH instability. The thin dotted lines passing through each panel set delimit the interfaces where the secondary instabilities trigger and the flow collapses into stratified turbulence.}
\label{R1500burst}
\end{figure}

An {\color{black}{interesting}} feature of figure \ref{Fig5a} we did not {\color{black}{discuss}} yet, is the robustness in the local modification of the velocity profile. {\color{black}{The region where fluctuations decelerate the mean flow is localised in
the vertical direction}}. As the energy increases, this thin layer shrinks further and hence, the velocity gradients strengthen. To have a clearer view, we present snapshots of velocity and buoyancy profiles in figure \ref{R1500burst} at both extrema of a bursting episode (denoted by the stars in figure \ref{Fig5b}). It is evident from the representation of density fluctuations in figure \ref{Fig6a} that domains of quiescent flow, surrounding the area of lowest horizontal velocity, slightly enhance the stratification profile. The two layers encompassing this region are both subject to the strongest vertical shear and therefore, correspond to the lowest values of $\Ri_z$ in figure \ref{Fig6b}. At the peak of energy, the stratification becomes unstable (with overturning) at these coordinates and the flow breaks down into a localized layer of chaotic motions. Strikingly, these disordered motion remain confined within a fixed interval (delimited by the thin dotted lines), before returning back to the original layered configuration displayed in figure \ref{Fig6a}. This {\color{black}{cyclical}} dynamics is well depicted through the Hovm\"oller diagram in figure \ref{Fig5a}.

\begin{figure}
\centering

\subfloat[]{
\includegraphics*[width=.8\textwidth]{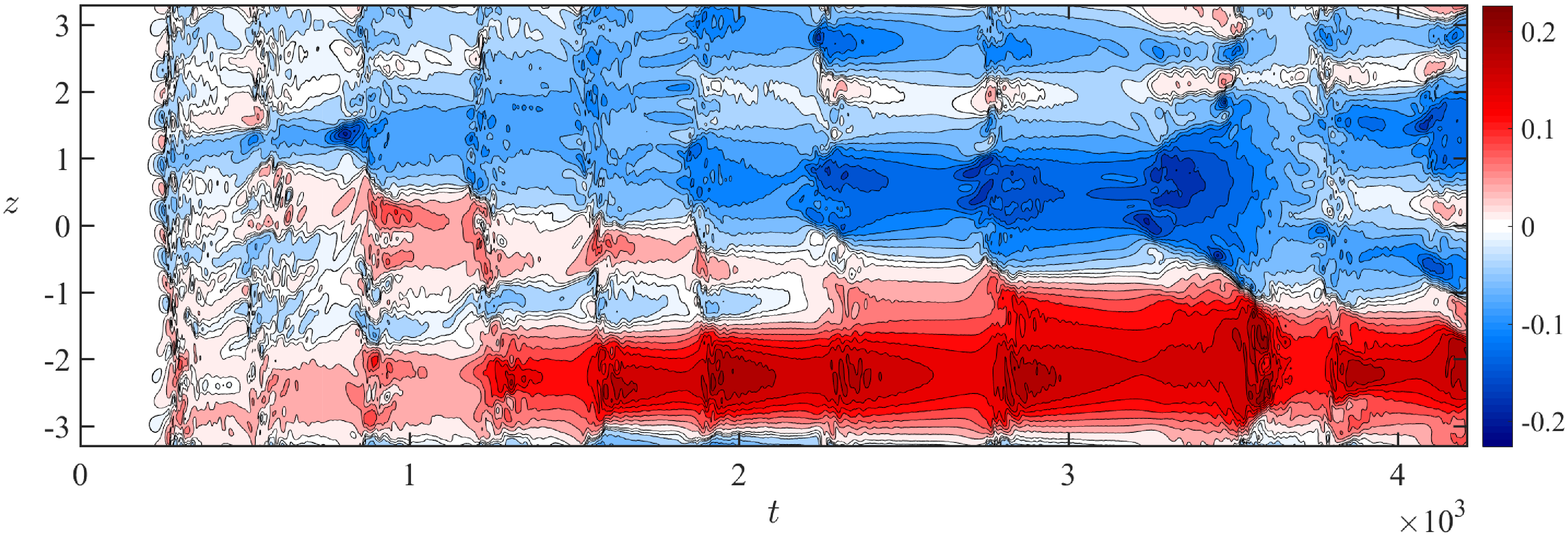}\label{Fig7a}} \\ \vspace*{-1em}
\subfloat[]{
\includegraphics*[width=.8\textwidth]{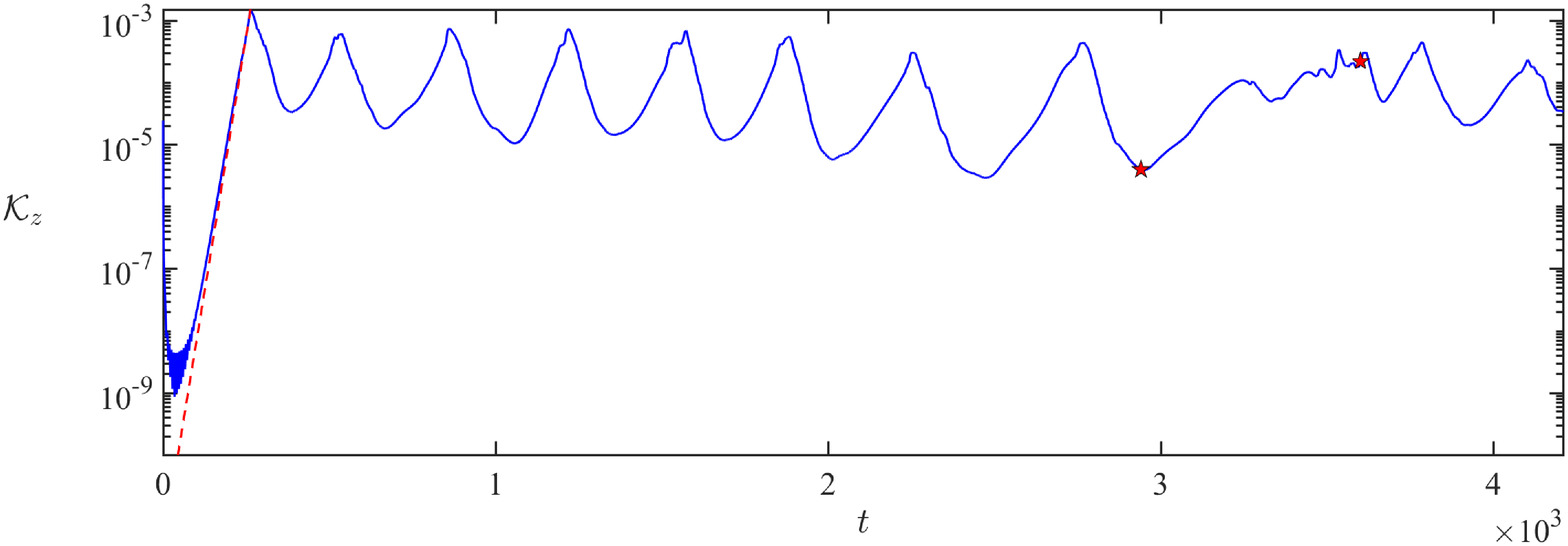}\label{Fig7b}} \\ \vspace*{-1em}
\subfloat[]{
\includegraphics*[width=.8\textwidth]{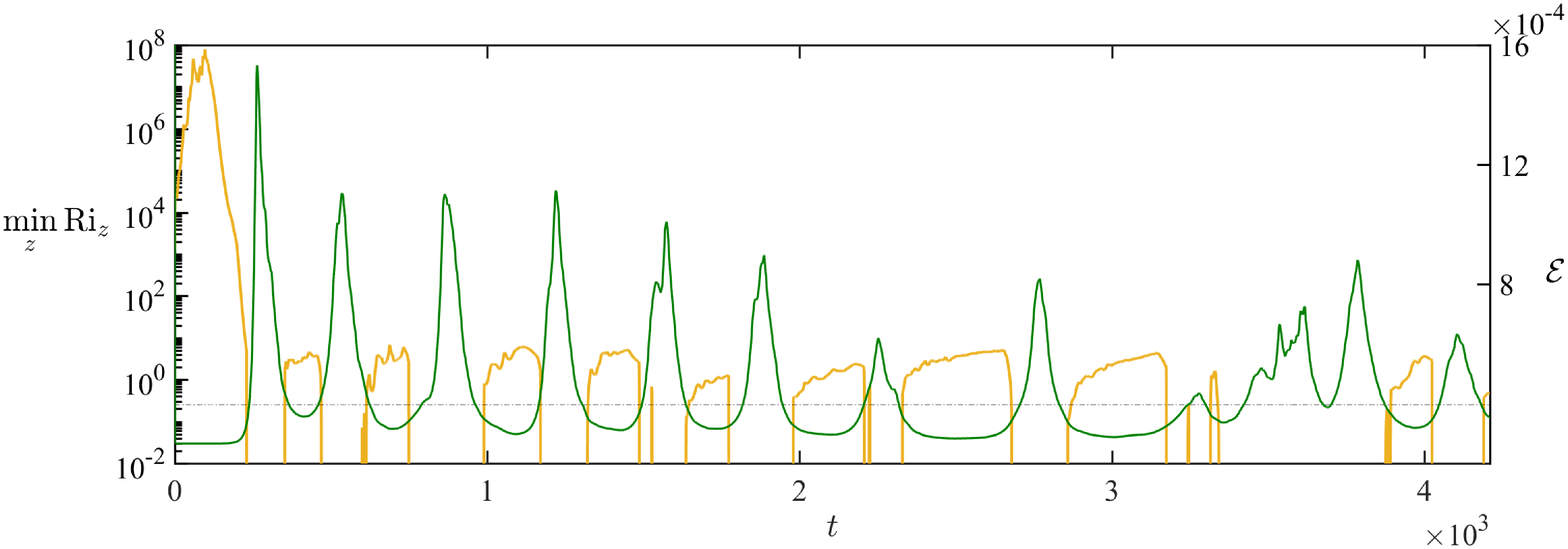}\label{Fig7c}}

\caption{DNS at $Re=5000$. Same panels as in figure \ref{R1500map}. The red markers in the representation of vertical kinetic energy indicate the time coordinates used in the snapshots of figure \ref{R5000snap}.}
\label{R5000map}
\end{figure}

\begin{figure}
\centering

\subfloat[]{
\includegraphics*[width=.35\textwidth]{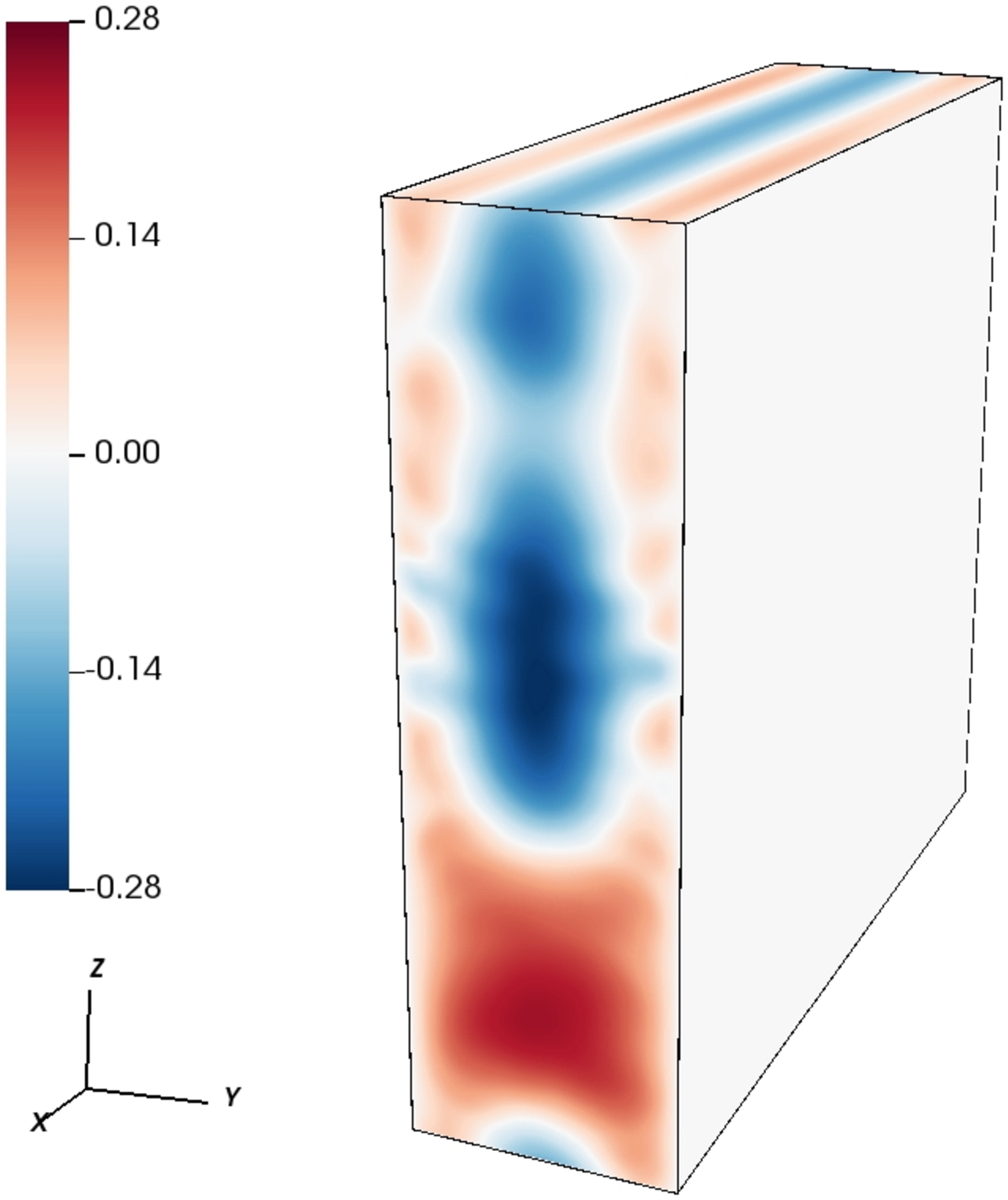}
\includegraphics*[width=.35\textwidth]{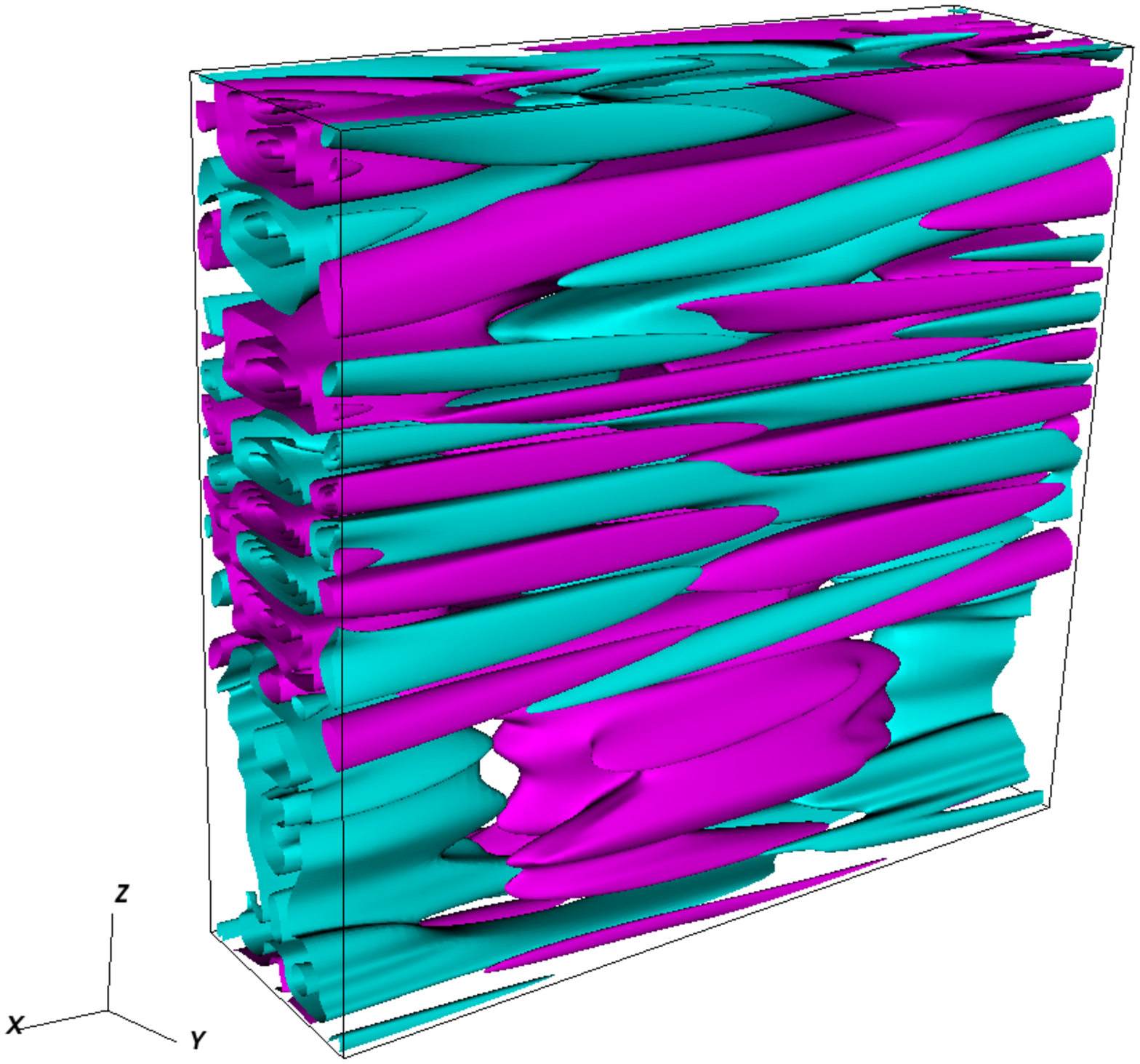}
\label{Fig8a}
} \\
\subfloat[]{
\includegraphics*[width=.35\textwidth]{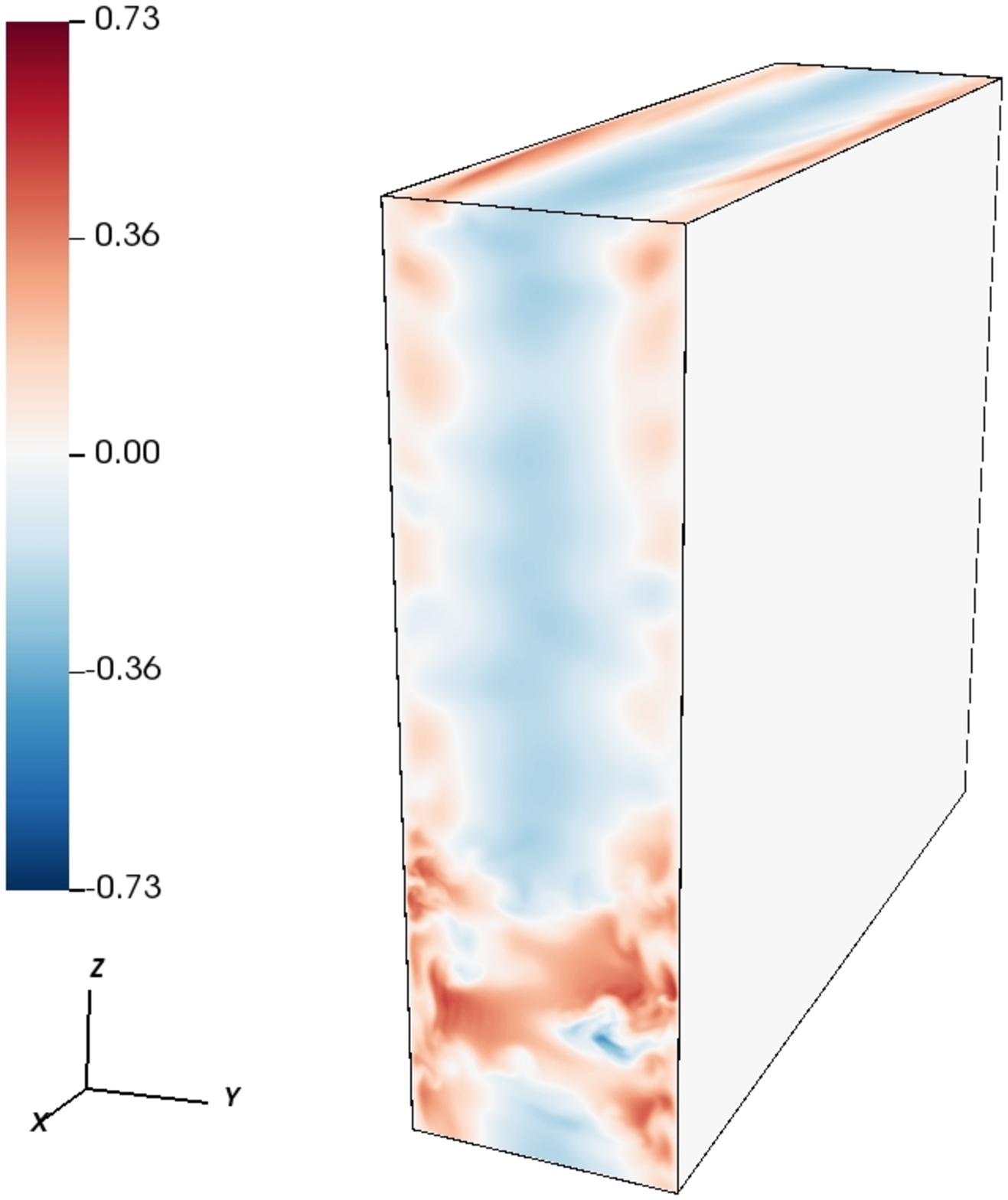}
\includegraphics*[width=.35\textwidth]{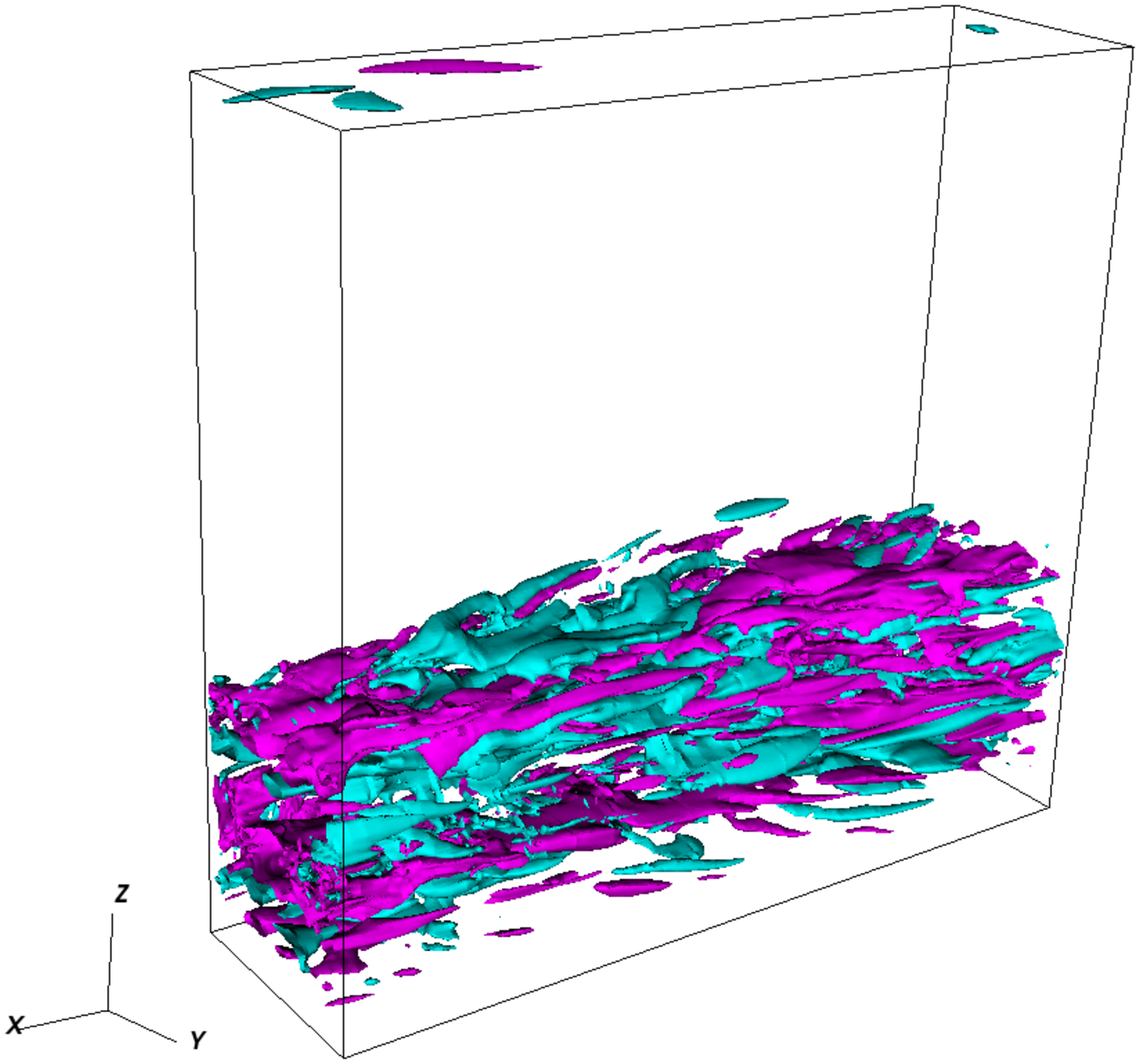}
\label{Fig8b}
}

\caption{Snapshots of the numerical computation at $Re=5000$, computed at (a) $t=2940$ and (b) $t=3600$. Left panels of each column display the streamwise velocity fluctuation $u-U_0$, whereas right panels represent the associated iso-contours of wall-normal velocity. Positive (negative) iso-contours are represented in light purple (light blue) regardless of their magnitude.}
\label{R5000snap}
\end{figure}

Increasing further the Reynolds number up to $Re=5000$, we present in figure \ref{R5000map} the {\color{black}{same set of data as in figure \ref{R1500map}}}. One instantaneous observation that one can make about the first panel is the complexity of the Hovm\"oller diagram in the large $Re$ configuration. The interfaces are indeed no longer formed in an organized way, as previously, but they follow instead a coarse-graining dynamics of intermittent patterns. {\color{black}{Regions of positive/negative velocity fluctuations emerge spontaneously
due to bursting overturns which can produce localised layers of turbulence}}. In the meantime, we still observe through the simulation a dominant mode slowly emerging, at the size of the domain. This leads us to think that the resonance of harmonics with the fundamental mode dominates the other interactions. Once again, the vertical kinetic energy displays a series of {\color{black}{cyclic}} episodes while the local Richardson numbers are now almost always below the threshold of instability, or even imaginary. It is thus not surprising to notice additional regions of intermittency in the data, such as in figure \ref{Fig6b}, as turbulent states now largely influence the global dynamics. As we did for the case of $Re=1500$, we present in figure \ref{R5000snap} {\color{black}{two snapshots}} of horizontal velocity, taken at the times of lowest and largest energy peaks in a bursting episode. We notice one large region subject to shear instabilities, further collapsing into highly chaotic motions. Although the bursting events are now more forceful than previously, the turbulence still {\color{black}{dies out}} before starting over a new cycle. This dynamics seems to persist over the simulations and we expect the system to behave similarly as the vertical extension is modified, as soon as resonances are permitted within the numerical domain. To demonstrate this behaviour, we compute in both panels of figure \ref{R5000snap} a visualization of the velocity field at $t=2940$ first, {\color{black}{where the flow is assumed to be mostly laminar, and then at $t = 3600$, where a localized turbulent band is observed. This band is of course reminiscent of the turbulent bands observed during the transition of unstratified shear flows. As already pointed, the link between these inclined turbulent bands and exact invariant solutions of the Navier--Stokes equation was addressed by \cite{Reetz} in the case without stratification. Indeed, starting from the periodic Nagata equilibrium \citep{Nagata}, \cite{Reetz} show that it is possible to search for localized flow pattern by windowing the initial pattern in a process already used and fully described in \cite{GibsonBrand}. Then imposing periodicity of the tilted domain in which the solution is searched, \cite{Reetz} show that it is possible to find alternating laminar and turbulent bands as observed in experiments and numerical simulations. The characteristics of these bands correspond (at least in the case of the Couette flow) to the observed pattern with a wavelength ranging from $40$ to $60$ times the gap. \cite{Reetz} show also that the inclination angle of the bands is found between $16.5^{\circ}$ and $26.1^{\circ}$ which is fully consistent with observations. Today, it is not known if the horizontal turbulent layers that we observe in our numerical simulations of the stratified Poiseuille flow have any connection with the localized bands of the unstratified Poiseuille flow. One first difference is the value of the selected wavelength: in our case, the modulation affects the entire flow domain as it emerges from the interaction of close neighbor modes contrary to the unstratified case where the pattern possesses its own periodicity. Another obvious difference between both flow patterns is the inclination angle of the turbulent regions which is zero in the case with stratification and seem above $16^{\circ}$ without. Note that this comparison was documented by \cite{Liu} where it is shown that (at least for the Couette flow) in the low-Reynolds and low Richardson number regime, the spatial intermittency is associated to oblique turbulent bands which are qualitatively similar to those seen in unstratified plane Couette flow. On the contrary, in the high Reynolds and high Richardson number regime, quasi-horizontal flow structures resembling the turbulent–laminar layers as observed commonly in stratified flows take place. \cite{Liu} suggest moreover that the classical Miles–Howard stability criterion (expressed by a Richardson number less than $1/4$) is associated with a change in the selection of the flow structures for a Schmidt number equal to $1$, which is our case here.}}

\section{Discussion}
\label{Sec7}

{\color{black}{In this article}}, we have presented a series of results on the plane Poiseuille flow with stable vertical stratification. One of the main reasons we studied this configuration is, as it was originally done in the unstratified limit, to study the transition from a laminar flow, subject to a linear instability, towards the onset and development of turbulence. First, we found that this system loses its vertical homogeneity as a consequence of an interaction of modes, whenever the numerical domain is elongated enough to allow unstable harmonics of the fundamental mode to exist. The result of this symmetry-breaking mechanism is the spontaneous formation of vertical velocity gradients in the initially homogeneous Poiseuille profile. This additional shear induces secondary instabilities othat we believe to be of KH type, as well as overturning events, confined in regions where the gradient Richardson number crosses the well-known critical threshold \citep{H61,M61}. Once this instability saturates and collapses into stratified turbulence, energy declines and the system returns to the state it formerly was in, before the onset of instability. This quasi-periodic dynamics of bursting episodes, enclosed within finite layers, is a remarkable finding that, to the best of our knowledge, {\color{black}{has never been observed in the literature on stratified fluids}}. However, it is not clear at this stage how the mean flow modulation will manifest itself in the limit of extremely elongated domain in the vertical direction, where a large number of unstable modes can coexist even close to the threshold. It also remains to verify whether such modulation can also occur in the streamwise direction if multiple horizontal wavelengths can fit into the numerical domain.

An important feature of this flow is, as mentioned, the spontaneous layering and the onset of turbulence, due to subsidiary shear instabilities. Moreover, these turbulent motions are encompassed between thin layers of finite size, delimited by interfaces where the Richardson number is the lowest. It is further observed that such regions are surrounded by a flow displaying the same meandering patterns as originally noticed by \cite{G21}. As the Reynolds number is increased, intermittency takes place and we are no longer able to distinguish an organized formation of turbulent layers. In this context, configurations with several {\color{black}{sets}} of layers (all undergoing a series of bursting events) {\color{black}{are allowed}} to exist. These observations tend to highlight the presence of a similar homoclinic snaking mechanism {\color{black}{that}} some dynamical systems are subject to \citep{BK07}. In recent articles, close connections between the field of pattern formation and the development of coherent structures, related to the layering of the flow, have been extensively studied \citep{LCK17}, notably in the context of unstratified plane Couette flow \citep{SGB10,GS16}. These works demonstrated the existence of invariant solutions, restricted to one spatial dimension, using Newton-like algorithms to perform a parameter continuation. Eventually, we could consider a similar treatment in our framework, although computation of vertically-localized solutions for such configuration is numerically expensive and requires therefore a better understanding of the weakly nonlinear dynamics.

To explain the symmetry loss in the nonlinear regime, we decomposed the disturbances as a superposition of linear modes, while setting ourselves in the vicinity of the instability threshold. This allowed us to extract an explicit form for the Reynolds stress, assuming the domain contains a finite number of harmonics. Although this approach is useful to understand the mechanism responsible for the onset of a secondary bifurcation and for the emergence of space-time modulations, it still adopts arguments from linear theory, which are no longer strictly valid in this regime. To avoid such impediment, a next step would be to investigate the weakly nonlinear analysis of the system, with finite amplitudes, shortly after the deviation from linear instability. With such a study, we would be able to isolate the term responsible for the correction of the mean flow and derive a set of nonlinear equations to describe its dynamics (in the form of a modified Stuart--Landau equation). Building upon the seminal works on the nonlinear Poiseuille configuration \citep{S60,SS71}, by means of a multiple scaling approach, we are confident that explicit results can be retrieved in our context.

\section*{Acknowledgments}

Centre de Calcul Intensif d’Aix-Marseille is acknowledged for granting access to its high-performance computing resources. This work was granted access to the HPC resources of IDRIS under the allocation 2022-A0120407543 made by GENCI.
The authors would like to acknowledge Yohann Duguet for interesting discussions and suggesting relevant references.

\section*{Funding}

This work received support from the French government under the France 2030 investment plan, as part of the Initiative d'Excellence d'Aix-Marseille Universit\'e - A{*}MIDEX (AMX-19-IET-010).

\section*{Declaration of interests}

The authors report no conflict of interest.

\nocite{*}
\bibliographystyle{jfm}
\bibliography{Biblio}

\end{document}